\documentclass[10pt,journal,compsoc]{IEEEtran}
\pdfoutput=1

\usepackage[nocompress]{cite}
\usepackage{listings}
\usepackage[pdftex]{graphicx,colortbl}
\DeclareGraphicsExtensions{.pdf,.jpg,.png}
\usepackage[cmex10]{amsmath}
\usepackage{amssymb}
\usepackage{array}
\usepackage[caption=false,font=footnotesize,labelfont=sf,textfont=sf]{subfig}
\usepackage{booktabs}
\usepackage[usenames]{xcolor}
\definecolor{lightgreen}{rgb}{0.56, 0.93, 0.56}
\usepackage{siunitx}
\usepackage{url}
\usepackage{enumitem}

\usepackage{xspace}

\usepackage{algorithm}
\usepackage{algorithmicx}
\usepackage[noend]{algpseudocode}
\usepackage{enumitem}
\algnotext{EndFor}
\algnotext{EndIf}
\algnotext{EndWhile}
\algnotext{EndProcedure}

\newcommand\verysmallfont{\fontsize{8}{8}\selectfont}
\newcommand\veryestsmallfont{\fontsize{7}{7}\selectfont}

\def\naive{{na\"{\i}ve}\xspace}

\newcommand\newtilde{\raise.17ex\hbox{$\scriptstyle\sim$}}

\newcommand{\vsf}[0]{\vspace{-0pt}}
\newcommand{\rb}[1]{\raisebox{1.5ex}[0pt]{#1}}
\newcommand{\comment}[1]{}

\newcommand{\Pulp}{\textsc{PuLP}\xspace}
\newcommand{\DGL}{\textsc{DGL}\xspace}

\begin{document}

\title{Distributed Graph Layout for Scalable Small-world Network Analysis}
%
%
%
%

\author{George~M.~Slota,~\IEEEmembership{Member,~IEEE,}
        Sivasankaran~Rajamanickam,~\IEEEmembership{Member,~IEEE,}~and~Kamesh~Madduri%
\IEEEcompsocitemizethanks{\IEEEcompsocthanksitem G. M. Slota is with the Computer Science Department, Rensselaer Polytechnic Institute, 317 Lally Building, Troy, NY, 12180, USA, USA. E-mail: slotag@rpi.edu %
\IEEEcompsocthanksitem S. Rajamanickam is with the Center for Computing Research at Sandia National Laboratories, P.O. Box 5800, MS 1320, Albuquerque, NM, 87123, USA. E-mail: srajama@sandia.gov%
\IEEEcompsocthanksitem K. Madduri is with the Department of Computer Science and Engineering, The Pennsylvania State University, 343E IST Building, University Park, PA, 16802, USA. E-mail: madduri@cse.psu.edu
}
}

\maketitle

\begin{abstract}
  The in-memory graph layout or organization has a considerable impact on the time and energy efficiency of distributed memory graph computations. It affects memory locality, inter-task load balance, communication time, and overall memory utilization. Graph layout could refer to partitioning or replication of vertex and edge arrays, selective replication of data structures that hold meta-data, and reordering vertex and edge identifiers. In this work, we present \DGL, a fast, parallel, and memory-efficient distributed graph layout strategy that is specifically designed for small-world networks (low-diameter graphs with skewed vertex degree distributions). Label propagation-based partitioning and a scalable BFS-based ordering are the main steps in the layout strategy.  We show that the \DGL layout can significantly improve end-to-end performance of five challenging graph analytics workloads: PageRank, a parallel subgraph enumeration program, tuned implementations of breadth-first search and single-source shortest paths, and RDF3X-MPI, a distributed SPARQL query processing engine. Using these benchmarks, we additionally offer a comprehensive analysis on how graph layout affects the performance of graph analytics with variable computation and communication characteristics.
\end{abstract}


%

\section{Introduction}\label{s:intro}

%
%
%
%

Layouts of graphs and sparse matrices in distributed memory and shared memory have been well-studied for regular graphs that arise in the scientific computing domain. ``Layout'' in this instance refers to how vertices and edges are partitioned in distributed-memory and how the vertex identifiers are ordered in shared-memory. Recently, several new open-source distributed-memory graph processing frameworks have emerged into mainstream usage. These include GraphLab~\cite{graphLab} and its derivatives PowerGraph~\cite{PowerGraph} and PowerLyra~\cite{PowerLyra}, Giraph~\cite{Giraph}, Trinity~\cite{Trinity}, and PEGASUS~\cite{PEGASUS}, among others. The primary goal of these frameworks is to analyze real-world graphs such as web crawls and social networks, which tend to be low-diameter graphs with skewed vertex degree distributions. Most of these frameworks assume an initial topology-agnostic vertex and edge partitioning and ordering. With these frameworks and small-world irregular graphs in mind, this paper attempts to answer the following questions:

\begin{enumerate}[nolistsep, topsep=0pt, leftmargin=*]
  \item Will the layout of the graphs impact the performance of
  irregular, data-analytic algorithms and frameworks ?
  \item Can such a layout be computed in a scalable and efficient fashion
  to be applicable in graph analytics ?
  \item What kind of graph computations will be impacted by the
  graph layouts and how ?
\end{enumerate}

As has been observed, the impact of partitioning and ordering on irregular graph computations can be considerable~\cite{CongPaper, FrascaSC12, boman2D}. However, using traditional layout strategies based on graph/hypergraph partitioners and orderings for data layout of highly irregular small-world graphs may not be appropriate for the following reasons:

\begin{enumerate}[nolistsep, topsep=0pt, leftmargin=*]
  \item Traditional partitioners and even some ordering methods, for example nested dissection, are heavyweight tools that are expensive both in terms of memory usage and time. They are appropriate when followed by more expensive linear solvers or when the partitioning results can be used for multiple solves. In contrast, graph analytic workloads are constantly evolving and a typical analytic operation is typically cheaper than a linear solver.
  \item Previous ordering algorithms are designed for metrics appropriate for linear solvers such as minimizing a bandwidth \cite{CMalg} or minimizing the fill-in in a LU factorization\cite{AMDalg, METIScode}. In contrast, ordering methods that improve the layouts in a shared memory context for small-world graphs are needed.
  \item The performance of distributed-memory graph algorithms can be dependent on both local and global graph topology. Global topology affects the number of parallel phases and synchronization overhead, while local topological changes impact per-phase load balance. Optimizing for aggregate measures such as conductance or edge cut would ignore local topology changes and may not account for dynamic variations in per-phase execution.
\end{enumerate}

Graph computations on highly irregular graphs require a layout that depends on parallel partitioners and ordering methods that are highly scalable for very large graphs. Label propagation-based partitioners are shown to be useful for partitioning small-world graphs~\cite{kahip}. We utilize such partitioning algorithms (\Pulp\cite{pulp2}) to compute the distributed memory layout. Label propagation exploits the community structure inherent in many real small-world graphs to quickly partition even multi-billion edge networks. Label propagation also allows for optimization of various objectives under multiple constraints, which enables us to explore the impact of these objectives and constraints on total execution and communication times for our various test analytics. In addition, we also introduce a breadth-first search-based ordering that is more scalable than other ordering schemes and suitable for small-world graphs in the shared-memory layout. In case of distributed graph processing, we consider various partitioning-ordering possibilities, a simultaneous global partitioning and ordering of all vertices, and a local ordering of vertices after the partitioning phase. 

In short, we propose a ``distributed-memory graph layout'' based on vertex partitioning using label propagation and a BFS-based parallel ordering strategy. The proposed \DGL (Distributed Graph Layout) is a fast, memory-efficient, and scalable graph layout strategy.  We demonstrate the new \DGL layout scheme is about 10-12$\times$ times faster to compute than METIS partitioning~\cite{METIScode}, and about 2.3$\times$ faster to compute than Reverse Cuthill-McKee (RCM)-based orderings. 


We demonstrate the impact of \DGL and present detailed analysis on the end-to-end performance of distinct graph analytic workloads. The graph analysis routines include subgraph counting, breadth-first search (BFS), single-source shortest paths (SSSP), resource description framework (RDF) queries, and PageRank. The five algorithms were chosen to be representative of the diversity in modern graph analytics. We chose a recent algorithm for subgraph counting~\cite{fascia} which is a randomized parallel algorithm to generate approximate counts of tree-structured subgraphs. Although recent related work~\cite{PetriniBFS, PetriniSSSP} primarily looks at strong scaling of BFS and related computations on massive synthetic Graph 500 networks, our work examines the subgraph counting algorithm, an analytic that is computationally very different from BFS. However, we also do an in-depth evaluation of BFS and SSSP performance. The fourth benchmark evaluates a distributed-memory implementation of the popular RDF store RDF-3X~\cite{RDF3X}. Our final included algorithm is a highly scalable implementation of PageRank~\cite{ipdps16}, which is a popular and more computationally-intensive implementation than BFS for benchmarking performance of frameworks and systems. 

We use the \emph{end-to-end} graph analysis times for partitioning-ordering-workload in both single-threaded (MPI) and multi-threaded (MPI+OpenMP) distributed programming models. We also consider computation and communication times of the analytic separately, in order to better isolate the effects of partitioning and ordering on performance. We primarily consider \emph{real-world} rather than synthetic graphs in our study. We use tuned implementations, all developed by us, in order to ensure consistency. We also analyze trade-offs between partitioning quality on computational load balance and communication overhead for several large real-world networks. The following is a summary of the key observations and findings from this workload analysis.

\begin{enumerate}[nolistsep, topsep=0pt, leftmargin=*]
  \item A comprehensive study of the performance of the five analytics with several partitioning-ordering combinations.
  \item Our \DGL ordering strategy is about 2$\times$ faster than RCM, and our \Pulp partitioning strategy is about 10$\times$ faster than METIS.
  \item We show that \DGL layout improves subgraph counting performance by 1.28$\times$ in comparison to random partitioning. Partitioning with \Pulp would enable end-to-end processing (partitioning \& computation) of the counts of ten vertex subgraphs on the 2 billion edge Twitter graph to complete in under fifteen minutes on 16 nodes of \emph{Blue Waters}.
  \item \DGL layout improves the communication time of BFS and SSSP by 1.48$\times$ and 1.43$\times$ in comparison to random partitioning.
  \item An informed topology-aware graph layout benefits external memory computations as well, improving the performance of RDF3X-MPI, our distributed-memory implementation of the popular RDF store RDF-3X~\cite{RDF3X}.
  \item The total computation time of PageRank can be accelerated by about 5$\times$ with a locality-optimizing ordering such as \DGL.
  \item A cross-analytics comparison reveals new and interesting trade-offs of communication time, load balance, and memory utilization for various graphs. 
\end{enumerate}

We finally mention that \DGL is not limited to the MPI processing models considered in this work, and can therefore be utilized as a preprocessing step while running under other graph engines and parallel execution environments.

\section{Distributed Graph Layout}

In this section, we discuss the distributed graph layout using label
propagation-based partitioning and BFS-based ordering methods. We define a
distributed graph layout as the pair of \emph{partitioning$\times$ordering}. The
partitioning part of the layout  affects the number of parallel phases and
synchronization overhead in a graph computation. It is important to balance the
computation in different parallel phases as well as minimize the communication
overhead. We explore trade-offs in work and memory balance and communication
minimization between tasks with different partitioning strategies. Work
performed and memory utilization per-task roughly correlates with the number of
vertices and adjacent edges stored on each task. The communication requirements
roughly correlates with the number of inter-task edges, or edge cut resulting
from partitioning. The ordering part of the layout affects the per-phase
computation time in graph computations. We ideally want to increase intra-node
memory access locality to reduce cache misses and improve execution times. In
order to be practical the \emph{partitioning$\times$ordering} pair must be
computed in parallel, scalable fashion.

\subsection{Partitioning}

We utilize three partitioners in this work. We use a random partitioning to establish a baseline for benchmarking, which randomly assigns part assignments to each vertex. We use the well-known METIS~\cite{METIScode} partitioner as a representation of the state-of-the-art. 

We also utilize the \Pulp partitioner, which is specifically optimized to partition the small-world graphs we are considering in this work. We consider both single constraint and multi-constraint partitioning scenarios, where we either balance partitions for vertices or for both vertices and edges. We attempt to minimize total edge cut for both \Pulp and METIS. Additionally, for \Pulp, we also attempt to balance communication among parts by minimizing the maximal number of cut edges coming out of any single part.


The \Pulp partitioner is based off of the community detection label propagation algorithm~\cite{RAK07}. Label propagation methods are attractive as they have low computational overhead, low memory utilization, are easy to parallelize, and demonstrate scaling to graphs with billions of vertices.

An overview of the basic label propagation algorithm is as follows: Initially, each vertex in a graph is initialized to having a unique label. Iteratively, each vertex then examines all of its neighbors' labels then updates to its label the label that appears most frequently among its neighbors, with ties broken randomly. The loop over all vertices can be parallelized without any explicit synchronizations or locking with minimal effect on solution quality~\cite{pulp2}. We continue to loop over all vertices until no labels are updated, or, more commonly, after some number of iterations of the outermost \emph{While} loop (usually 10 or fewer iterations is sufficient).

\begin{algorithm}[htb]
  \verysmallfont  
  \begin{algorithmic}[htb]
    \State Initialize $p$ parts
    \State Execute degree-weighted label propagation.
    \For{$k_1$ iterations}
      \State Balance parts for vertex constraint.
      \State Refine parts to minimize edge cut.
    \EndFor
    \For{$k_2$ iterations}
      \State Balance parts to satisfy edge constraint
      \State ~~and minimize max per-part cut.
      \State Refine parts to minimize edge cut.
    \EndFor
  \end{algorithmic}
  \caption{\Pulp Multi-Constraint Multi-Objective Algorithm Overview}
  \label{alg:pulp}
\end{algorithm}

\Pulp's subroutines essentially use variants label propagation that limit the number of possible labels to the number of desired parts and impose additional weighting criteria to create balanced partitions. This weighted form of label propagation is utilized in two separate stages during execution of \Pulp. Algorithm~\ref{alg:pulp} gives a very broad overview of the \Pulp multi-constraint (vertices and edge per part) multi-objective (minimize total edge cut and maximum edge cut per part) algorithm that demonstrates these two stages. After initialization, we first utilize weighted label propagation in $k_1$ alternating stages to balance the initial parts for our vertex constraint and then refining to minimize the total edge cut. Next, we perform $k_2$ alternating stages of balancing for our edge constraint while minimizing the secondary objective of max per-part cut and then again refining to minimize the total edge cut. In prior work~\cite{pulp2}, we describe the algorithm in considerably greater detail and demonstrate the approach's effectiveness in terms of cut quality and runtime with respect to other traditional partitioners. However, it is critical to show that such label propagation-based partitionings are not only easy to compute, but that they also improve the end-to-end runtimes of graph analytic applications. With \DGL, we are able to utilize such a partitioner in the layout strategy and demonstrate its applicability for the first time.

\subsection{Ordering}

For a distributed graph computation, a good graph partitioning will reduce inter-node communication cost. The goal of on-node vertex ordering is to increase locality of intra-node memory references, and thereby reduce intra-node computation time. This is done by relabeling vertex identifiers so that consecutive accesses of per-vertex specific information occur with greater spatial and temporal locality. As many graph computations access per-vertex data based on adjacencies, and per-vertex data is commonly stored in a flat array, minimizing the numeric difference between the vertex identifiers of adjacent vertex pairs can greatly improve access locality and therefore cache utilization.

Reverse Cuthill-McKee (RCM) is a commonly-used vertex ordering strategy in sparse matrix and graph applications. We propose a BFS-based ordering (see Algorithm~\ref{alg:bfs_ordering}) which can be considered an approximation to RCM. It avoids the costly sorting step used in RCM where it tries to order the nodes with the same parent in terms of the degree. Recently, a similar ordering was proposed for improving the matrix-vector multiply time and bandwidth reduction~\cite{parallelRCM}. The primary focus of that approach was to arrive at parallel orderings to improve the linear solver time. Our focus is to improve the graph computations' end-to-end time. 

We demonstrate our approach in Algorithm~\ref{alg:bfs_ordering}. We randomly choose a minimal-degree vertex as the $root$ and perform a standard BFS routine, tracking visitation status with $visited$ and the current level with $level$. We add vertices to level sets $L$ when they are visited, as with RCM. We avoid explicit sorting by assuming that each $L_{0\cdots Max_{level}}$, where $Max_{level}$ is the maximum BFS level, is mostly sorted in the order of decreasing vertex degree, as there is a higher likelihood of encountering high-degree vertices sooner in any given level for most real world graphs. We assign new labels using an incrementing value of $n$ by starting with the vertices in the highest level and working backwards to those in the lowest level. As we will show in the next section, this approach performs better than both random and RCM orderings in applications that have a high number of irregular memory accesses. As with RCM in~\cite{parallelRCM}, Algorithm~\ref{alg:bfs_ordering} can be straightforwardly parallelized.

\begin{algorithm}[htb]
	\verysmallfont
	\begin{algorithmic}[htb]
		\State $V_{id} \gets$ DGL-order($G(V, E)$)
		\ForAll{$v \in V$}
		  \State $V_{id}(v) \gets v$
		\EndFor
		\State $level \gets 0$
    \State $root \gets$ SelectRoot()
    \State $Q \gets root$
		\State $Visited(1\cdots |V|) \gets$ \textbf{false}
		\While{$Q \neq \varnothing$}
		  \ForAll{$v \in Q$}
		    \State Insert $v$ into $L_{level}$
        \ForAll{$\langle v, u\rangle \in E$}
		      \If{$Visited(u) = $ \textbf{false}}
		        \State $Visited(u) \gets $ \textbf{true}
		        \State Insert $u$ into $Q$
		      \EndIf
		    \EndFor
		  \EndFor
      \State $level \gets level + 1$
		\EndWhile
    \State $Max_{level} \gets level$
		\State $n \gets 0$
		\For{$i = Max_{level} \cdots 1$}
		  \For{$j = 1\cdots |L_i|$}
		    \State $V_{id}(L_i(j)) \gets n$
		    \State $n \gets n + 1$
		  \EndFor
		\EndFor
	\end{algorithmic}
	\caption{\DGL BFS-based vertex ordering algorithm.}
	\label{alg:bfs_ordering}
\end{algorithm}

We compare the performance of our ordering method and RCM to a random ordering, where vertices are shuffled into a random order. Another common ordering strategy is using the ``natural'' ordering, which is using the original vertex assignments as given in the graph data file. We avoid comparison to the natural ordering (and associated vertex block partitioning) for a couple of reasons. The quality of the natural layout for graphs retrieved from a database can be highly variable. E.g. a natural layout for a web graph that's based on crawling methodology might give considerably better performance than the layout for a social graph that's based off of user ID (we observed a spread of both 2$\times$ speedup and slowdown for a natural ordering vs. random ordering on our PageRank test; similarly, we observed consistent variance in vertex block vs. random partitioning). Additionally, a ``natural'' layout for certain graphs is not necessarily well-defined. For example, the user IDs for a social network might be considerably larger than the number of vertices in the graph. To create and efficiently store a graph in-memory, these values need to be mapped to vertex identifiers $0\cdots |V|-1)$. This mapping can be done by \emph{compressing} the user IDs and retaining their original order, or through a more efficient first-come-first-served mapping where IDs are mapped to vertex identifiers as they are encountered in the data file. The method used would be application-specific. Because the extensive number of tests we performed limited the number of graphs we could use in our experiments, we wished to eliminate any potential impacts due to the above graph creation methodologies for the sake of uniformity in our comparative results. Hence why we view a random ordering to be the best baseline for relative comparison.

With the five partitioning methods (random, METIS \{single constraint, single objective and multiple constraint, single objective\} and \Pulp \{multiple constraint, single objective and multiple constraint, multiple objective\}) and three ordering methods (random, RCM, and DGL) we evaluate all the combinations of \emph{partitioning$\times$ordering} pairs and demonstrate that the DGL layout with \Pulp partitioner and \DGL-based ordering performs the best in irregular graph computations.

\section{Parallel Graph Computations}

In this section, we will give an overview of the five distributed graph analytics used during our experimental analysis of the impact of partitioning and ordering on analytic performance. In an attempt to best understand the general effects of varying partitioning and ordering on the performance, the graph analytics were selected as to represent a wide range of execution characteristics. The test suite includes an implementation which is relatively computation-heavy, PageRank, algorithms which are relatively more communication-heavy, breadth-first search and single source shortest paths, an algorithm which is both very computation and communication intensive, color-coding subgraph counting, as well as an algorithm whose performance is dependent on the sizes of the $n$-hop neighborhoods of each partition, distributed query processing of Resource Description Framework stores.

\subsection{Distributed PageRank}


Our distributed PageRank uses an MPI+OpenMP approach and an $\frac{|V|}{p}$ partitioning, with each of $p$ MPI tasks calculating the counts for an equivalent portion of the $|V|$ vertices in the graph $G$. With one MPI task per node, we then use thread parallelism while updating the counts of owned vertices. With the exception of the single MPI communication call on each iteration, all per-task work can be done in parallel. Updates are passed among neighbors using an MPI all-to-all exchange. In practice, this specific implementation has been observed to be very efficient and scalable, giving per-iteration costs of less than a few seconds for networks of over 100 billion edges while running on 256 compute nodes. The specific technical details of the implementation are omitted, but please see~\cite{ipdps16} for a more in-depth discussion.

\subsection{Subgraph Counting}

Subgraph counting is a computationally challenging task, with the \naive approach scaling as $O(n^k)$, where $n$ is the number of vertices in a graph and $k$ the number of vertices in the subgraph being counted. The best known exact algorithm~\cite{EG04} improves the exponent by a factor of $\frac{\alpha}{3}$, where $\alpha$ is the exponent for fast matrix multiplication. Because of these extremely high execution time bounds, recent work has focused on approximation algorithms. One such approach for counting \emph{tree-structured} subgraphs utilizes the color-coding technique of Alon et al.~\cite{AYZ95}.


In prior work, we developed a fast parallel implementations of color-coding subgraph counting in both shared-memory and distributed-memory environments~\cite{fascia}. The distributed version of our approach uses several optimizations, including fully partitioning and compressing the memory-intensive dynamic programming table to decrease memory requirements across all tasks, further compressing the table during communication to reduce the total transfer volume, and using all-to-all exchanges in lieu of broadcasts to reduce communication times. These optimizations demonstrate good scaling and enable us to count subgraphs of 10 and 11 vertices on billion-edge networks in minutes on a modest number of 16 nodes. For space consideration, we omit a detailed description of our implementation. Instead, please refer to~\cite{fascia} for an in-depth discussion of the stages and execution of the algorithm.


\subsection{SSSP and BFS}

We also assess the performance impact of layout on tuned implementations for parallel breadth-first search (BFS) and single-source shortest paths (SSSP) computation in this paper. Our parallel BFS approach can take advantage of both 1D and 2D graph distributions~\cite{BM11b, BM13, boman2D}. We use a 1D distribution in this work, as it is easier to correlate communication time with edge cut after partitioning with a 1D distribution. Recent BFS and SSSP implementations use a 1D partitioning and direction-optimizing search~\cite{hybridBFS} for work-efficient and highly scalable execution on Graph 500 test instances. For an overview of the current state-of-the-art in performance optimizations for these routines, we refer the reader to~\cite{PetriniSSSP, PetriniBFS}.

We use an optimized parallel implementation~\cite{PM14b} of the $\Delta$-stepping algorithm~\cite{MS03} for parallel SSSP in this paper. Each BFS iteration and $\Delta$-stepping phase is comprised of three main steps: local discovery, all-to-all exchange, and local update. To aid adjacency queries, we use a distributed compressed sparse row representation for a graph. The distance array is also partitioned and distributed along with the distributed vertices (for $\Delta$-stepping). In the local discovery step, both algorithms expand their frontiers by listing all corresponding adjacencies and their proposed distance based on vertices in a queue of recently-visited vertices (for BFS) or in a current bucket (for $\Delta$-stepping). Note that BFS visits each reachable vertex only once while $\Delta$-stepping may visit each reachable vertex multiple times before it is settled. 

Once all vertices in the queue are processed or the current bucket is empty (with no more vertex reinsertions), all $p$ tasks exchange vertices in these generated lists to make them local to the owner tasks. This step is the same for both BFS and $\Delta$-stepping, and uses an all-to-all collective communication routine. At the end of each BFS iteration and $\Delta$-stepping phase, each task locally updates the distance of its own vertices using the exchanged information. The update in BFS is only on unvisited vertices, while $\Delta$-stepping updates all vertices whose distances can be decreased. Thus, the $\Delta$-stepping algorithm performs more computation and has a higher communication complexity. 

Since our goal is to analyze and evaluate the effect of graph partitioning and vertex reordering, we have not yet implemented all the optimizations in~\cite{PetriniSSSP, PetriniBFS}. However, our approach has three new optimizations: (i) A semi-sort of vertex adjacencies based on weights is used prior to execution of the algorithm. (ii) Memory-optimized queues are used to represent the bucket data structure. This decreases the algorithm memory requirement, while slightly increasing the running time. (iii) An array of all local unique adjacencies is created and locally used to track tentative distance of  adjacencies. This array improves efficiency by filtering out unnecessary requests to be added in the new frontiers.

\subsection{Distributed RDF Stores and SPARQL Query Processing}

Resource Description Framework (RDF)~\cite{RDFPrimer04} is a popular data format for storing web data sets. Informally, the RDF format specifies typed relationships between entities, and the basic record in an RDF data set is a \emph{triple}. There are a growing number of publicly-available RDF data sets that contain billions of triples. Thus, database methodologies for storing these RDF data sets, also called triple stores, are becoming popular. We have developed a distributed MPI-based implementation of an open-source triple store called RDF-3X~\cite{RDF3X}. Our distributed RDF store is called RDF3X-MPI~\cite{rdf3xmpi}. 

An alternate approach to viewing an RDF data set is as a directed graph with edge types. RDF data sets can be queried using a language called SPARQL. We extend the distributed RDF store methodology of RDF-3X to the SPARQL querying phase as well. Thus our RDF3X-MPI tool has two phases, a load phase and a query phase. In the load phase, the given triple data set is partitioned into several independent files, one per task, and each task then constructs an index for helping answering SPARQL queries. It is possible to parallelize some query evaluation in a purely data-parallel manner (i.e., with no communication between tasks), provided there is sufficient replication of triples among partitions. Formally, if the triple partitions satisfy an \emph{$n$-hop guarantee}, then SPARQL queries in which all pairs of join variables are at distance of less than $n$ hops from each other can be solved without any inter-task communication~\cite{HAR11}. So the role of graph partitioning in this application is to reorder vertices such that the number of triples that are replicated between tasks after applying an $n$-hop guarantee are minimized. If the number of triples that are replicated is reduced, then the database indexes are smaller, making them potentially faster to query. For this application, we study the impact of partitioning on the number of replicated triples. A smaller value of replication is desired, and further, smaller index sizes should translate to faster query times.

\section{Experimental Setup}

We evaluate performance of our new partitioning and ordering strategy \DGL and the graph analytics workload on a collection of nine large-scale low diameter graphs, listed in Table~\ref{table:graphs}. LiveJournal, Orkut, and Twitter (follower network) are crawls of online social networks obtained from the SNAP Database and the Max Planck Institute for Software Systems~\cite{SNAP,twitter}. uk-2005 and sk-2005 are crawls of the United Kingdom (.uk) and Slovakian (.sk) domains performed in 2005 using UbiCrawler and downloaded from the University of Florida Sparse Matrix Collection~\cite{webgraph,UbiCrawler,UF}. WebBase is similarly a crawl obtained in 2001 by the Stanford WebBase crawler. We created the BSBM and LUBM graphs from RDF data sets generated using the Berlin SPARQL benchmark~\cite{BSBM} and Lehigh University Benchmark~\cite{LUBM} generators. DBpedia was created from RDF triples extracted from Wikipedia~\cite{DBpedia}.

The Orkut graph is undirected and the remaining graphs are directed. For the web and social graphs, we preprocessed the graphs before executing PageRank, BFS, SSSP, and subgraph counting. Specifically, we removed all degree-0 vertices, multi-edges, and extracted the largest (weakly) connected component. Further, edge directivity was ignored when partitioning the graphs using \Pulp and METIS and reordering with RCM and \DGL. Table~\ref{table:graphs} lists the sizes of these nine graphs after preprocessing. 

\begin{table}[htb]
	\verysmallfont
	\centering
	\caption{\verysmallfont Test graph characteristics \emph{after} preprocessing. Graphs belong to three categories, OSN: Online social networks, WWW: Web crawl, RDF: graphs constructed from RDF data. \# Vertices ($n$), \# Edges ($m$), average ($d_{avg}$) and max \ ($d_{max}$) vertex degrees, and approximate diameter ($\widetilde{D}$) are listed. $B = \times 10^9$, $M = \times 10^6$, $K=\times 10^3$.}\vsf
	\tabcolsep=0.1cm
	\begin{tabular}{llrrrrrc} \\
		\toprule
		Network & Category & 
		\multicolumn{1}{c}{$n$} & \multicolumn{1}{c}{$m$} & 
		\multicolumn{1}{c}{$d_{avg}$}  & \multicolumn{1}{c}{$d_{max}$} & 
		\multicolumn{1}{c}{$\widetilde{D}$} & \multicolumn{1}{c}{Source}\\ 
		\midrule
		LiveJournal & OSN & 4.8~M &   42~M &   18 &    39~K &   21 & \cite{lj}\\
		Orkut       & OSN & 3.1~M &  117~M &   76 &    33~K &    9 & \cite{orkut}\\
		Twitter     & OSN &  44~M &  2.0~B &   37 &   750~K &   36 & \cite{twitter}\\
		uk-2005     & WWW &  39~M &  781~M &   40 &   1.8~M &   21 & \cite{webgraph}\\
		WebBase     & WWW & 113~M &  844~M &   15 &   816~K &  376 & \cite{webgraph}\\
		sk-2005     & WWW &  44~M &  1.6~B &   73 &    15~M &  308 & \cite{webgraph}\\
		BSBM        & RDF &  16~M &   67~M &  8.6 &   3.6~M &    7 & \cite{BSBM}\\
		LUBM        & RDF &  33~M &  133~M &  8.1 &    11~M &    6 & \cite{LUBM}\\
		DBpedia     & RDF &  62~M &  190~M &  6.1 &   7.3~M &    7 & \cite{DBpedia}\\
		\bottomrule
	\end{tabular}
	\label{table:graphs}
\end{table}


The scalability studies for subgraph counting, BFS, SSSP, and RDF query processing were done primarily on \emph{Blue Waters}, a large petascale supercomputer at the National Center for Supercomputing Applications (NCSA). Each XE compute node of \emph{Blue Waters} is a dual-socket system with 64~GB main memory and AMD 6276 Interlagos processors at 2.3 GHz. The system uses a Cray Gemini 3D torus interconnect. We built our programs with the GNU C++ compiler (version 4.8.2), using OpenMP for multithreading and the \texttt{-O3} optimization parameter during compilation. For the pre-processing phases of DGL (partitioning and reordering) and some scalability runs, we utilized \emph{Compton}, a testbed cluster. \emph{Compton} has a dual socket setup with Intel Xeon E5-2670 (Sandy Bridge) CPUs at 2.60~GHz and 64~GB main memory. Due to the large memory requirements of partitioning with METIS, we also had to use the large memory nodes on \emph{Carver} at NERSC for partitioning the larger networks (Twitter, uk-2005, Webbase, and sk-2005). Carver's large memory nodes have 1024~GB main memory and four Intel Xeon X7550 ("Nehalem-EX") CPUs at 2.00~GHz. We performed k-way partitioning with METIS using version 5.1.0.


\section{Results and Discussion}

\subsection{\DGL Performance Evaluation}

We first evaluate our \DGL label propagation-based partitioning methodology, \Pulp, against METIS partitioning by examining total running time for generating 16 and 64 partitions. We consider two versions of both \Pulp and METIS. For \Pulp, we have an implementation that has \emph{both} maximal vertex and edge balance constraints and minimizes \emph{both} total edge cut and maximal per-part edge cut. We consider this our baseline implementation, and label it in figures as \Pulp-MM (\Pulp multi-objective multi-constraint). We also have a dual constraint version that only attempts to minimize the total edge cut, which we call \Pulp-M. Similarly for METIS, the dual constraint single objective version is termed METIS-M, while the single constraint (vertex balance) and single-objective version is termed simply as METIS. METIS-M and \Pulp-M are solving the same problem. For our constraints, we fix the maximal vertex imbalance ratio at 1.10 and the edge imbalance ratio at 1.50. The results will show that the multi-constraint, multi-objective mode of \Pulp-MM can be important for irregular graph computations.

Table~\ref{table:part_time} shows the partitioning time of \Pulp-MM along with METIS-M running on \emph{Compton}. Due to METIS's large memory requirements (close to 500GB for Twitter), only LiveJournal, Orkut, and the RDF graphs could be partitioned on \emph{Compton}. The larger web graphs and Twitter were all partitioned on a large memory node of Carver. We also report the relative speedup of \Pulp to METIS. We omit time comparison to ParMETIS, as the only graphs it was able to successfully partition on any system were LiveJournal and Orkut. Further, ParMETIS's speedups relative to METIS for those two instances were minimal (less than 2$\times$ with 16-way parallelism). From Table~\ref{table:part_time} we observe considerable speedup for \Pulp, with a geometric mean speedup of 12.4$\times$ for 16 parts and 10.1$\times$ for 64 parts.

\begin{table}[htb]
	\veryestsmallfont
	\centering
	\caption{\verysmallfont \Pulp-MM and METIS-M partitioning time with 16-way and 64-way partitioning. \Pulp-MM uses multi-constraint multi-objective partitioning. METIS-M uses multi-constraint single-objective partitioning.}\vsf
	\tabcolsep=0.1cm
	\begin{tabular}{lrrrrrr} \\
		\toprule
		& \multicolumn{3}{c}{16-way partitioning} & \multicolumn{3}{c}{64-way partitioning}\\
		\rb{Network}  & METIS-M & \Pulp-MM & & METIS-M & \Pulp-MM & \\
		& time (s)  & time (s) & \rb{Speedup} & time (s) & time (s) & \rb{Speedup}\\
		\midrule
		LiveJournal &    75 & 7.4 &  10$\times$ &    74 & 7.3 &  10$\times$\\
		Orkut       &   156 &  10 &  16$\times$ &   197 &  13 &  15$\times$\\
		Twitter     & 12348 & 530 &  23$\times$ & 12484 & 565 &  22$\times$\\
		uk-2005     &   255 &  15 &  17$\times$ &   353 &  80 & 4.4$\times$\\
		WebBase     &   539 &  39 &  14$\times$ &   551 &  42 &  13$\times$\\
		sk-2005     &   465 &  39 &  12$\times$ &   514 &  65 & 7.9$\times$\\
		BSBM        &   348 &  28 &  12$\times$ &   395 &  32 &  12$\times$\\
		LUBM        &   707 &  88 & 8.0$\times$ &   966 & 123 & 7.9$\times$\\
		DBpedia     &   898 & 133 & 6.8$\times$ &  1001 & 133 & 7.5$\times$\\  
		\bottomrule
	\end{tabular}
	\label{table:part_time}
\end{table}

\begin{table}[htb]
	\veryestsmallfont
	\centering
	\caption{\verysmallfont Average partitioning characteristics across all graphs. Geometric mean of vertex balance $V_{max}$, edge balance $E_{max}$, improvement over random partitioning for edge cut ratio $EC$ and max per-part edge cut $EC_{max}$, and the mean improvement (decrease) in the average total number of connected components for all parts (\#CCs) are shown. The best values for each of the last three columns are in bold font.}\vsf
	\tabcolsep=0.15cm
	\begin{tabular}{lrr|rrr|rrr}
    \toprule
    & & & \multicolumn{3}{c}{$EC(imp)$} & \multicolumn{3}{c}{$EC_{max}(imp)$}\\
    \rb{Partitioner} & $V_{max}$ & $E_{max}$ & avg & min & max & avg & min & max \\
    \midrule
		Random   & 1.15 & 1.70 & 1.00 & 1.00 & 1.00 & 1.00 & 1.00 & 1.00 \\
    METIS    & 1.10 & 3.88 & \textbf{7.71} & \textbf{1.5} & \textbf{107} & 2.39 & 0.25 & 63 \\
    METIS-M  & 1.10 & 1.50 & 4.40 & 1.02 & 41 & 2.16 & 0.77 & 22 \\
		\Pulp-M  & 1.10 & 1.50 & 5.50 & 1.17 & 64 & 2.10 & 0.54 & 23 \\
		\Pulp-MM & 1.10 & 1.50 & 5.00 & 1.19 & 63 & \textbf{3.18} & \textbf{2.54} & \textbf{204} \\
		\bottomrule
	\end{tabular}
	\label{table:part_qual}
\end{table}

The partitioning quality in terms of both vertex and edge balance constraints and edge cut and maximal per-part edge cut objectives for the different partitioners is shown in Table~\ref{table:part_qual} as geometric averages. We also note that aggregate measures don't fully capture the wide spread of results among different tests, so include \emph{min} and \emph{max} improvements for edge cut and max per-part cut as well. E.g., the improvement METIS has relative to random partitioning varies from 1.5$\times$ for 64-way partitioning of Twitter to 107$\times$ improvement for 16-way partitioning of uk-2005.

In terms of the total edge cut ($EC$), the single-constraint, single-objective METIS does the best, but it performs poorly in the maximum per-part edge cut ($EC_{max}$) and edge balance ($E_{max}$). \Pulp-MM also performs better than all the methods in the $EC_{max}$ metric without sacrificing a lot in $EC$ and still respecting the vertex balance and edge balance constraints. Also note the much larger $E_{max}$ of single constraint METIS. As we will demonstrate, this can have a considerably impact of execution time for the applications in our benchmarks. We bold the best values for each column for edge cut and max per-part cut, and note that METIS performs consistently best overall in the edge cut metric and \Pulp-MM performs best overall in the max per-part metric by a wide margin.



\begin{table}[htb]
	\veryestsmallfont
	\centering
	\caption{\verysmallfont \DGL serial reordering time with 16-way and 64-way partitioning.}\vsf
	\tabcolsep=0.1cm
	\begin{tabular}{lrrrrrr}
		\toprule
		& \multicolumn{3}{c}{16-way partitioning} & \multicolumn{3}{c}{64-way partitioning}\\
		\rb{Network}  & RCM  & \DGL & & RCM & \DGL & \\
		& time (s)  & time (s) & \rb{Speedup} & time (s) & time (s) & \rb{Speedup}\\
		\midrule
		LiveJournal & 2.3 & 1.0 & 2.3$\times$ & 2.3 & 1.0 & 2.3$\times$\\
		Orkut       & 3.9 & 1.9 & 2.1$\times$ & 3.9 & 1.9 & 2.1$\times$\\
		Twitter     &  50 &  24 & 2.1$\times$ &  61 &  29 & 2.1$\times$\\
		uk-2005     &  16 & 8.4 & 1.9$\times$ &  17 & 7.6 & 2.2$\times$\\
		Webbase     &  33 &  13 & 2.5$\times$ &  35 &  17 & 2.1$\times$\\
		sk-2005     &  24 &  11 & 2.2$\times$ &  23 &  11 & 2.1$\times$\\
		BSBM        & 5.1 & 2.3 & 2.2$\times$ & 4.7 & 2.3 & 2.0$\times$\\
		LUBM        & 5.7 & 1.7 & 3.4$\times$ & 5.7 & 1.7 & 3.4$\times$\\
		DBpedia     &  16 & 6.1 & 2.6$\times$ &  17 & 6.9 & 2.5$\times$\\
		\bottomrule
	\end{tabular}
	\label{table:reorder_time}
\end{table}

We additionally compare our \DGL vertex ordering strategy to RCM. Table~\ref{table:reorder_time} gives the average running times of both \DGL and RCM in serial across all three partitioning strategies for reordering the vertices within each partition. \DGL reordering results in a 2.3$\times$ average speedup compared to RCM for reordering both 16 and 64 parts. This reduction is due to the avoidance of explicit sorting required by RCM. There does not seem to be a large dependence of running times on the number of partitions, although with a greatly increased partition count for a fixed graph, it would be expected that running time decreases due to a lower diameter BFS search and overall increased cache utilization. Both these methods can be parallelized as \DGL can use a parallel BFS and RCM can be implemented using the parallel version \cite{parallelRCM}. However, their timings are insignificant in the end-to-end performance of complex analytics such as our subgraph counting benchmark.

\begin{table}[htb]
  \veryestsmallfont
  \centering
  \caption{\verysmallfont \DGL and \Pulp scaling to higher part counts. Execution times are in seconds.}\vsf
  \tabcolsep=0.1cm
  \begin{tabular}{lrrrrrr}
    \toprule
    & \multicolumn{2}{c}{256-way} & \multicolumn{2}{c}{512-way} & \multicolumn{2}{c}{1024-way}\\
    \rb{Network}  & \Pulp-MM  & \DGL & \Pulp-MM  & \DGL & \Pulp-MM  & \DGL \\
    \midrule
    LiveJournal &  10 & 1.2 &   18 & 1.2 &   30 & 1.2 \\
    Orkut       &  23 & 1.6 &   37 & 1.6 &   65 & 1.5 \\
    Twitter     & 910 &  42 & 1340 &  42 & 1560 &  41 \\
    uk-2005     & 109 & 6.6 &  161 & 6.7 &  252 & 6.7 \\
    Webbase     &  53 &  11 &  119 &  12 &  190 &  12 \\
    sk-2005     & 165 &  12 &  285 &  12 &  587 &  12 \\
    BSBM        &  55 & 1.6 &   75 & 1.6 &  127 & 1.6 \\
    LUBM        & 164 & 1.5 &  194 & 1.5 &  355 & 1.5 \\
    DBpedia     & 219 & 8.0 &  325 & 8.1 &  502 & 7.9 \\
    \bottomrule
  \end{tabular}
  \label{table:scaling_times}
\end{table}

We demonstrate that the \DGL and \Pulp strategies are also able to efficiently compute the layout for larger numbers of parts beyond 16 and 64. Table~\ref{table:scaling_times} gives the execution times of \DGL ordering and \Pulp partitioning when computing the layout of the various test graphs with 256, 512, and 1024 parts. We observe flat scaling of \DGL ordering with increasing part counts due to its intrinsic work efficiency and $O(n+m)$ expected execution time. We observe an increase in \Pulp times for higher part counts. This is due to \Pulp having a per-iteration workload of $O(np+m)$~\cite{pulp2}, where $p$ is the number of parts being computing. However, we still observe sub-linear scaling relative to $p$ for almost all test cases. Overall, we observe no intrinsic scalability bottlenecks of these methods at this higher scale.

We include one more table to demonstrate how our \DGL ordering strategy might improve cache performance of executing codes. To improve the performance of linear solvers, a common ordering metric to optimize for is graph bandwidth, which is the maximum integer distance between vertex identifiers for vertices that share a single neighbor. RCM is an effective means of bandwidth reduction for regular matrices. However, for small-world graphs, the bandwidth is usually going to be large, on the order of $d_{max}$, where $d_{max}$ is the maximal degree of any vertex in the graph. Comparing bandwidth measures between different orderings therefore won't show any global improvements in compaction for rows of much lesser degree vertices.

As such, we look at other metrics to give an indication of the possible cache efficiency in practice. Across the entire adjacency array, we measure how often edges listed in order also have identifiers within a single integer value of each other. This indicates that these edges would be neighboring nonzeros in the same row of an adjacency matrix. Co-located edges improve cache utilization of per-vertex information accesses, such as checking visitation status for BFS or PageRank value lookups. To quantify how many co-located vertex identifiers for the edges are in the adjacency list, we report two values. First, we report a ratio of how co-located all edges are, where a value of zero indicates that no edges are co-located and a value of one indicates that all edges are co-located. Second, we report a running ``cost'' as the sum of the distances, or gaps, between vertex identifiers in the adjacency list. We scale the distances by their $\log$, as a distance of one or close to it indicates that vertices are closely co-located and would have minimal cache cost for their subsequent accesses, and the cost difference between large and very large distances is minimal, since it's likely a new cache line would need to be loaded in both instances. Finding an ordering that minimizes this sum is referred to in the literature as the \emph{Minimum Logarithmic Gap Arrangement} problem, which is NP-hard~\cite{shingle}. In an attempt to give a graph-independent ratio, we further scale the $\log$ sum by a worst-case possible value of $m\log{n}$. The true dependence of cache utilization on distance would be architecture-specific, but the approximation of this cost gives enough insight for comparative purposes when examining ordering quality.

\begin{table*}[htb]
  \centering
  \caption{\verysmallfont Ordering performance for \DGL, RCM, and Random in terms co-location ratio (Co-loc.\ Ratio) and $\log$ sum of gap distances (Gap Sum Ratio) for 16-way and 64-way partitioning, averaged across the five different partitioning strategies.}\vsf
  \tabcolsep=0.25cm
  \begin{tabular}{lrr|rrr||rr|rrr}
    \toprule
    & \multicolumn{5}{c}{16-way partitioning} & \multicolumn{5}{c}{64-way partitioning}\\
    \rb{Network}  & \multicolumn{2}{c}{Co-loc.\ Ratio} & \multicolumn{3}{c}{Gap Sum Ratio} & \multicolumn{2}{c}{Co-loc.\ Ratio} & \multicolumn{3}{c}{Gap Sum Ratio} \\
    & \DGL & RCM & \DGL & RCM & Rand & \DGL & RCM & \DGL & RCM & Rand \\
    \midrule
    LiveJournal & 0.115 & 0.010 & 0.036 & 0.034 & 0.043 
                & 0.104 & 0.014 & 0.009 & 0.006 & 0.012 \\
    Orkut       & 0.028 & 0.001 & 0.046 & 0.057 & 0.054 
                & 0.021 & 0.001 & 0.010 & 0.020 & 0.011 \\
    Twitter     & 0.032 & 0.005 & 0.037 & 0.035 & 0.038 
                & 0.026 & 0.006 & 0.013 & 0.010 & 0.016 \\
    uk-2005     & 0.659 & 0.176 & 0.015 & 0.022 & 0.046 
                & 0.582 & 0.184 & 0.005 & 0.006 & 0.011 \\
    Webbase     & 0.562 & 0.162 & 0.020 & 0.039 & 0.050 
                & 0.519 & 0.172 & 0.005 & 0.006 & 0.011 \\
    sk-2005     & 0.613 & 0.149 & 0.018 & 0.026 & 0.050 
                & 0.689 & 0.167 & 0.002 & 0.004 & 0.013 \\
    BSBM        & 0.146 & 0.146 & 0.040 & 0.040 & 0.062 
                & 0.146 & 0.153 & 0.007 & 0.006 & 0.009 \\
    LUBM        & 0.105 & 0.105 & 0.026 & 0.026 & 0.045
                & 0.094 & 0.105 & 0.006 & 0.006 & 0.009 \\
    DBpedia     & 0.442 & 0.257 & 0.028 & 0.036 & 0.046 
                & 0.398 & 0.267 & 0.006 & 0.008 & 0.010 \\
    \midrule
    Overall & \textbf{0.298} & 0.112 & \textbf{0.030} & 0.034 & 0.048
            & \textbf{0.274} & 0.119 & \textbf{0.007} & 0.008 & 0.011 \\
    \bottomrule
  \end{tabular}
  \label{table:gaps}
\end{table*}

Table~\ref{table:gaps} gives both the co-location ratio (Co-loc.\ Ratio) as well as the $\log$ sum of gap distances ratios (Gap Sum Ratio) for all ordering combinations across all graphs for 16 and 64 parts. We report the geometric mean values across all five partitioning strategies (Random, METIS, METIS-M, \Pulp-M, \Pulp-MM). For co-location ratio, higher indicates better locality, while for the $\log$ gap sum ratio, lower indicates better locality. We omit reporting the co-location ratio for Random ordering in Table~\ref{table:gaps}, as all values are close to zero, a few orders of magnitudes less than RCM and \DGL. We observe nearly that \DGL ordering results in the best co-location ratio and lowest $\log$ gap sum ratio across almost all instances. The computational timings results we'll report next in our benchmarks will demonstrate that these measurements translate into real performance benefits across a wide range of graph analytics.

\subsection{PageRank Performance}

\begin{figure*}[htb]
  \centering
  \includegraphics[width=0.95\textwidth]{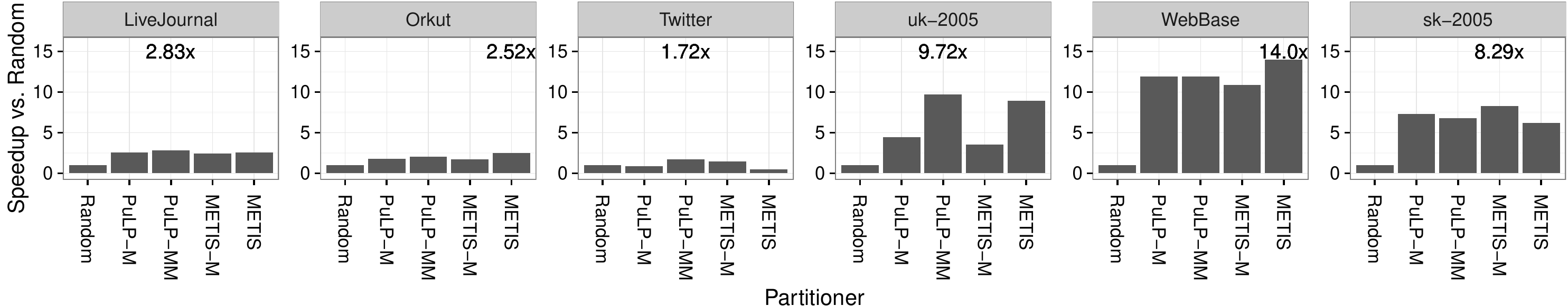}
  \includegraphics[width=0.95\textwidth]{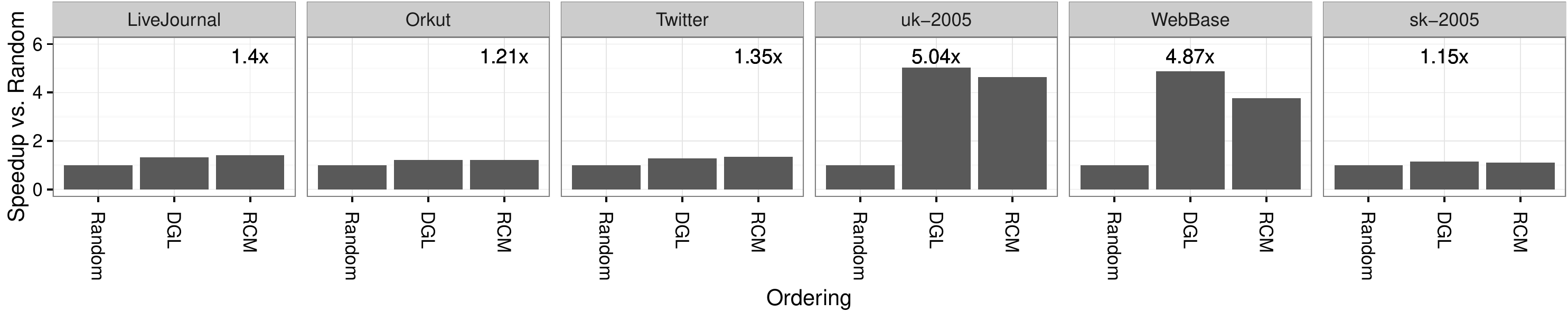}
  \caption{\verysmallfont Communication speedup of the PageRank implementation on 16 nodes with various partitioning options (top) and computation speedup of PageRank with various ordering strategies (bottom).}
  \label{fig:pr_perf}
\end{figure*}

\begin{table}[htb]
  \veryestsmallfont
  \centering
  \caption{\verysmallfont Speedups of various partitioning and ordering strategies versus random partitioning and random ordering for the PageRank counting benchmark.}\vsf
  \tabcolsep=0.1cm
  \begin{tabular}{lrrrr|rr}
    \toprule
    & \multicolumn{4}{c}{Partitioning} & \multicolumn{2}{c}{Ordering}\\
    \rb{Network}  & METIS & METIS-M & \Pulp-M & \Pulp-MM & RCM & \DGL \\
    \midrule
    LiveJournal & 2.560 & 2.453 & 2.561 & 2.832 & 1.404 & 1.325 \\
    Orkut       & 2.519 & 1.689 & 1.784 & 2.068 & 1.214 & 1.205 \\
    Twitter     & 1.454 & 1.459 & 1.871 & 1.716 & 1.346 & 1.292 \\
    uk-2005     & 8.913 & 3.518 & 4.427 & 9.725 & 4.641 & 5.039 \\
    WebBase     & 13.99 & 10.87 & 11.92 & 11.93 & 3.776 & 4.870 \\
    sk-2005     & 6.170 & 8.293 & 7.287 & 6.797 & 1.100 & 1.155 \\
    \midrule
    Overall & 3.621 & 3.525 & 3.395 & \textbf{4.465} & 1.881 & \textbf{1.970}\\
    \bottomrule
  \end{tabular}
  \label{table:pr_perf}
\end{table}

For our first set of experimental benchmarks results, we examine the effect of partitioning and ordering on a distributed PageRank implementation. We will first show the effect that different partitionings have on communication times, and then we will show the effect that orderings have on computation times. For these experiments we use the three social network graphs (LiveJournal, Orkut, and Twitter) as well as the three web crawls (uk-2005, WebBase, sk-2005). Figure~\ref{fig:pr_perf} (top) gives the speedups relative to random partitioning for METIS single and multiple constraint partitionings and \Pulp multiple constraint with single and multiple objective partitionings. Figure~\ref{fig:pr_perf} (bottom) gives the speedups relative to random ordering for \DGL and RCM. Table~\ref{table:pr_perf} gives the explicit speedup values and overall geometric means across the six test graphs. These value are for 20 iterations of PageRank executing on 16 nodes of \emph{Blue Waters}.

We observe that all partitionings offer considerable speedups relative to random. In general, the web crawls show even greater speedups than the social networks. This is due to the web crawls being greater in diameter and more separable than social networks, resulting in a decrease in the number of cut edges and subsequently greater performance improvements relative to random. Averaged across all six test graphs, \Pulp multiple constraint and multiple objective partitioning offers the greatest speedup. The performance benefit is due to the implementation's use of an iterative bulk synchronous model and moderately low required communication, so the improved communication balance resulting from \Pulp-MM's decrease in max per-part cut becomes apparent in the timings. 

Additionally, we note that both RCM and \DGL offer considerable speedups for total computation times relative to random ordering. On the uk-2005 and WebBase graphs, the speedups for \DGL are about 5$\times$. Again we observe that the social networks generally show less performance benefit relative to random, and this is again due to their lower diameter and small-world characteristics, which makes effective ordering more difficult. However, we still observe a consistent 20\%-40\% speedups with the improved orderings. Overall, \DGL gives a greater performance speedup over RCM by about 10\%, a result we expected based on our measurement of potential locality and cache performance as demonstrated in Table~\ref{table:gaps}.

\subsection{Subgraph Counting Performance}

\begin{figure*}[htb]
  \centering
  \includegraphics[width=0.95\textwidth]{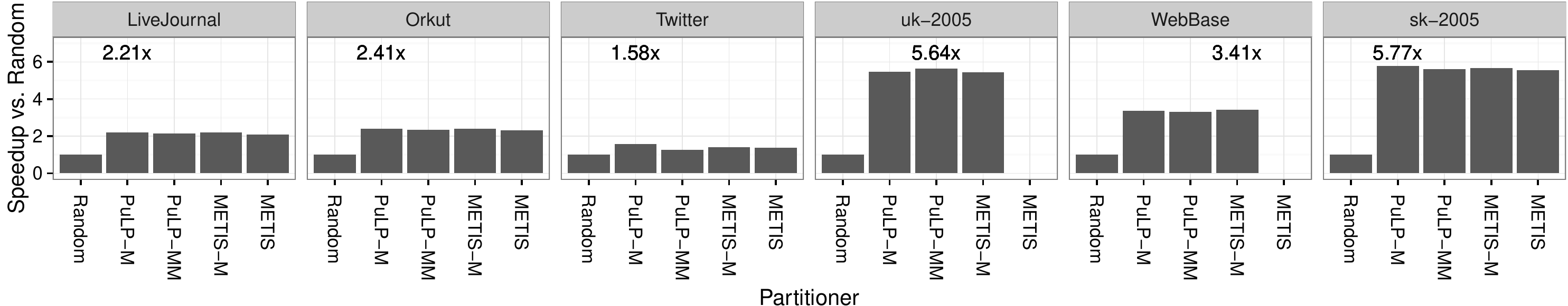}
  \includegraphics[width=0.95\textwidth]{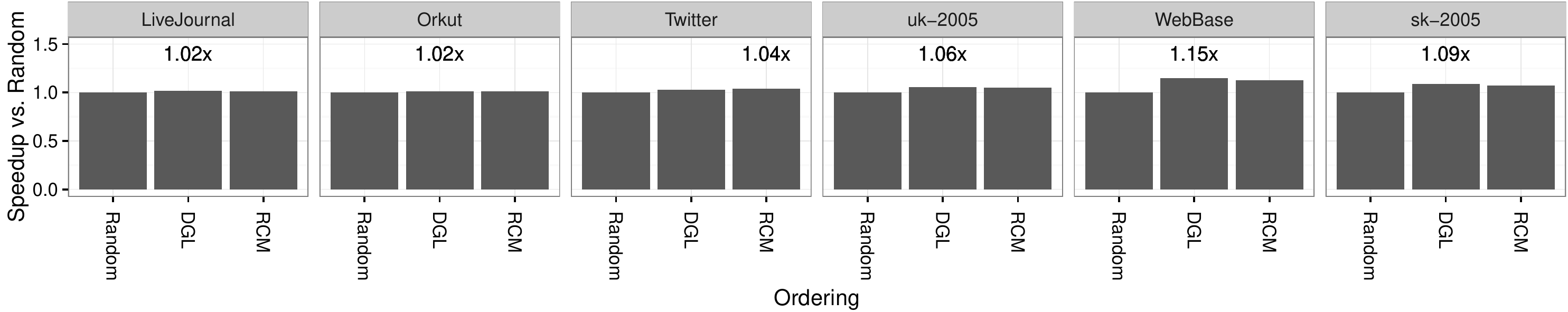}
  \includegraphics[width=0.95\textwidth]{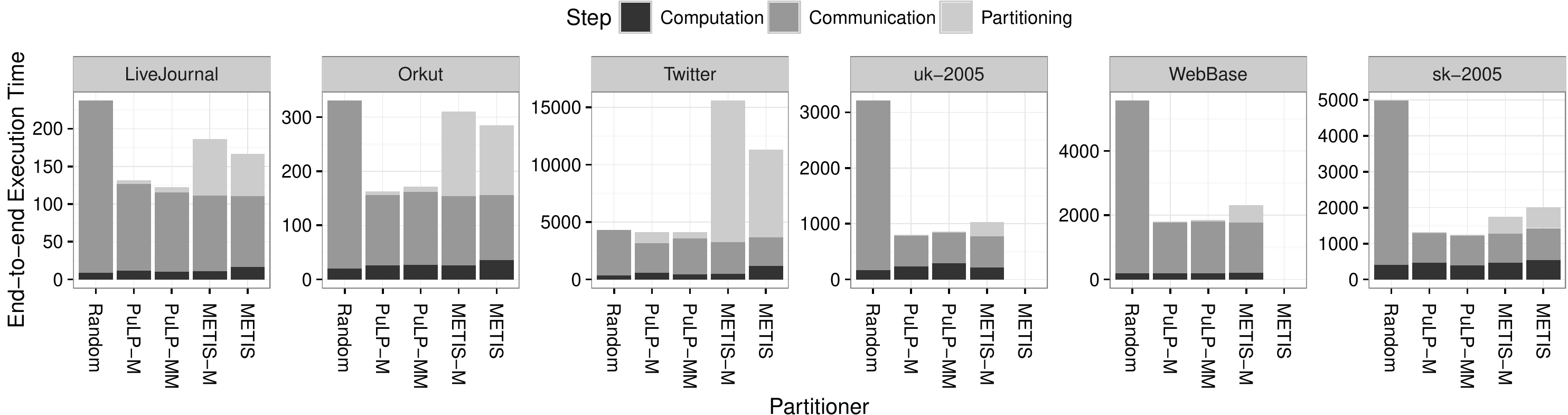}
  \caption{\verysmallfont Speedups achieved with subgraph counting for total communication time of the various partitioning strategies relative to random partitioning, all with random ordering. Additionally, the speedups for the RCM and \DGL orderings relative to random ordering with \Pulp multi objective partitioning. The bottom plot gives total end-to-end execution time in terms of the initial partitioning, total computation time, and total communication time, all in seconds.}
  \label{fig:fascia_perf}
\end{figure*}

\begin{table}[htb]
  \veryestsmallfont
  \centering
  \caption{\verysmallfont Speedups of various partitioning and ordering strategies versus random partitioning and random ordering for the subgraph counting benchmark.}\vsf
  \tabcolsep=0.1cm
  \begin{tabular}{lrrrr|rr}
    \toprule
    & \multicolumn{4}{c}{Partitioning} & \multicolumn{2}{c}{Ordering}\\
    \rb{Network}  & METIS & METIS-M & \Pulp-M & \Pulp-MM & RCM & \DGL \\
    \midrule
    LiveJournal & 2.099 & 2.202 & 2.211 & 2.150 & 1.009 & 1.020 \\
    Orkut       & 2.307 & 2.400 & 2.411 & 2.350 & 1.014 & 1.015 \\
    Twitter     & 1.378 & 1.399 & 1.580 & 1.271 & 1.041 & 1.029 \\
    uk-2005     & -     & 5.433 & 5.476 & 5.642 & 1.049 & 1.057 \\
    WebBase     & -     & 3.412 & 3.375 & 3.311 & 1.125 & 1.148 \\
    sk-2005     & 5.568 & 5.675 & 5.772 & 5.621 & 1.072 & 1.091 \\
    \midrule
    Overall     & - & 3.033 & \textbf{3.106} & 2.961 & 1.051 & \textbf{1.059} \\
    \bottomrule
  \end{tabular}
  \label{table:fascia_perf}
\end{table}

We next compare the impact of various partitioning and ordering strategies with regards to the running times of our subgraph counting implementation. We run on 16 node of \emph{Blue Waters}. We compare communication times resulting from each of the 5 partitioners with a fixed random ordering. We also compare the computation times resulting from the 3 ordering strategies with fixed \Pulp-MM partitioning. The speedups for each strategy on the 6 test graphs are given in Figure~\ref{fig:fascia_perf} and Table~\ref{table:fascia_perf}. We also look at total end-to-end execution time for the five partitioning strategies with random ordering in terms of total time spent in the communication, computation, and partitioning steps. Note that the results with single constraint METIS for the uk-2005 and WebBase graphs are absent. This is due to execution times taking longer than 24 hours for these instances.

Several trends can be observed in Figure~\ref{fig:fascia_perf}. The top subfigure gives the speedup of the communication phase of subgraph counting for each of the partitioning strategies relative to random partitioning. We again note considerable speedup for all partitioners. We note that the \Pulp methods give the best improvement for five out of the six tested graphs. Overall \Pulp-M gives the highest speedup overall. This implementation doesn't benefit as highly from the more communication-balanced \Pulp-MM partitioning due to the overall higher communication requirements (the Twitter graph requires compression and transfer of several terabytes of data in total for the $Count$ table exchanges between tasks) and lower overall synchronization cost relative to PageRank, so total edge cut is observed to have a greater effect in practice. This emphasizes the fact that a one-size-fits all solution is not optimal in practice, and implementation knowledge is required to extract the best performance for any given running application when utilizing a layout strategy.


The middle subfigure of Figure~\ref{fig:fascia_perf} plots the speedup relative to random ordering for the \DGL and RCM reordering strategies with \Pulp-MM partitioning. Overall, we note about a 6\% improvement for \DGL and 5\% improvement for RCM ordering relative to random. These improvements are much lower than PageRank's improvement due to considerably more information stored per-vertex in the stored counts table, so greater cache locality has less of an effect in preventing re-accesses to main memory; however, we note even a modest 5\%-6\% consistent improvement can be noteworthy in this instance. On processors with larger cache, this relative improvement would be expected to increase.



Finally, the bottom subfigure of Figure~\ref{fig:fascia_perf} shows the total end-to-end execution time in seconds for initial partitioning plus running of the subgraph counting application. We further split subgraph counting into the sum of time spent in each of its computation and communication phases. We observe that our partitioning and ordering strategies result in the fastest end-to-end running times for all test instances. The time spent for partitioning is considerable relative to execution time for METIS, as is the extra communication costs that result with random partitioning. The additional partitioning time cost for METIS might be amortized in practice by re-using the same partitions for subsequent analysis, but we note that \textbf{\Pulp partitioning shows an immediate decrease in total end-to-end time after a single analytic run}.


\subsection{Execution Timelines}

\begin{figure*}[htb]
  \centering
  \includegraphics[width=0.24\textwidth]{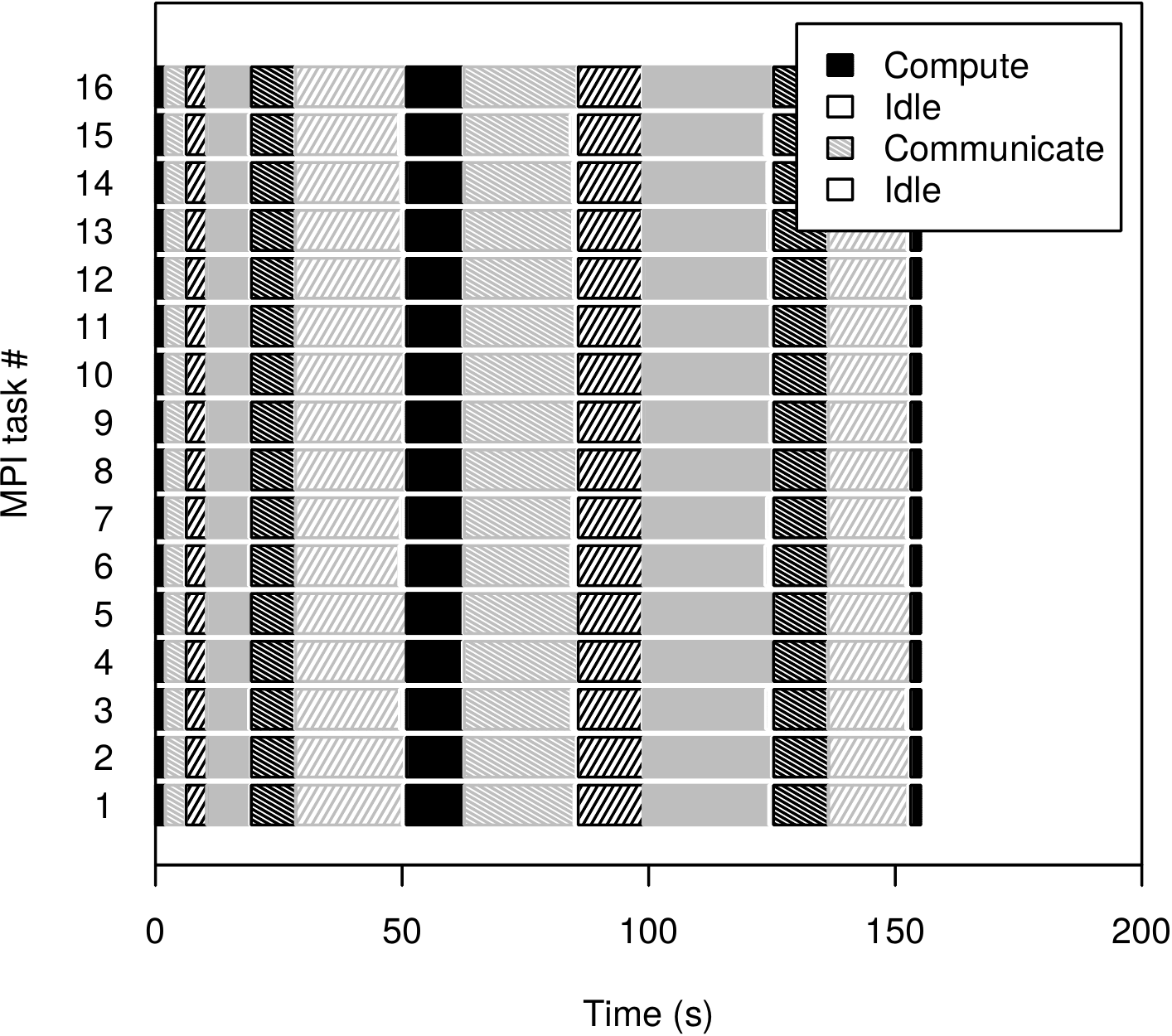}
  \includegraphics[width=0.24\textwidth]{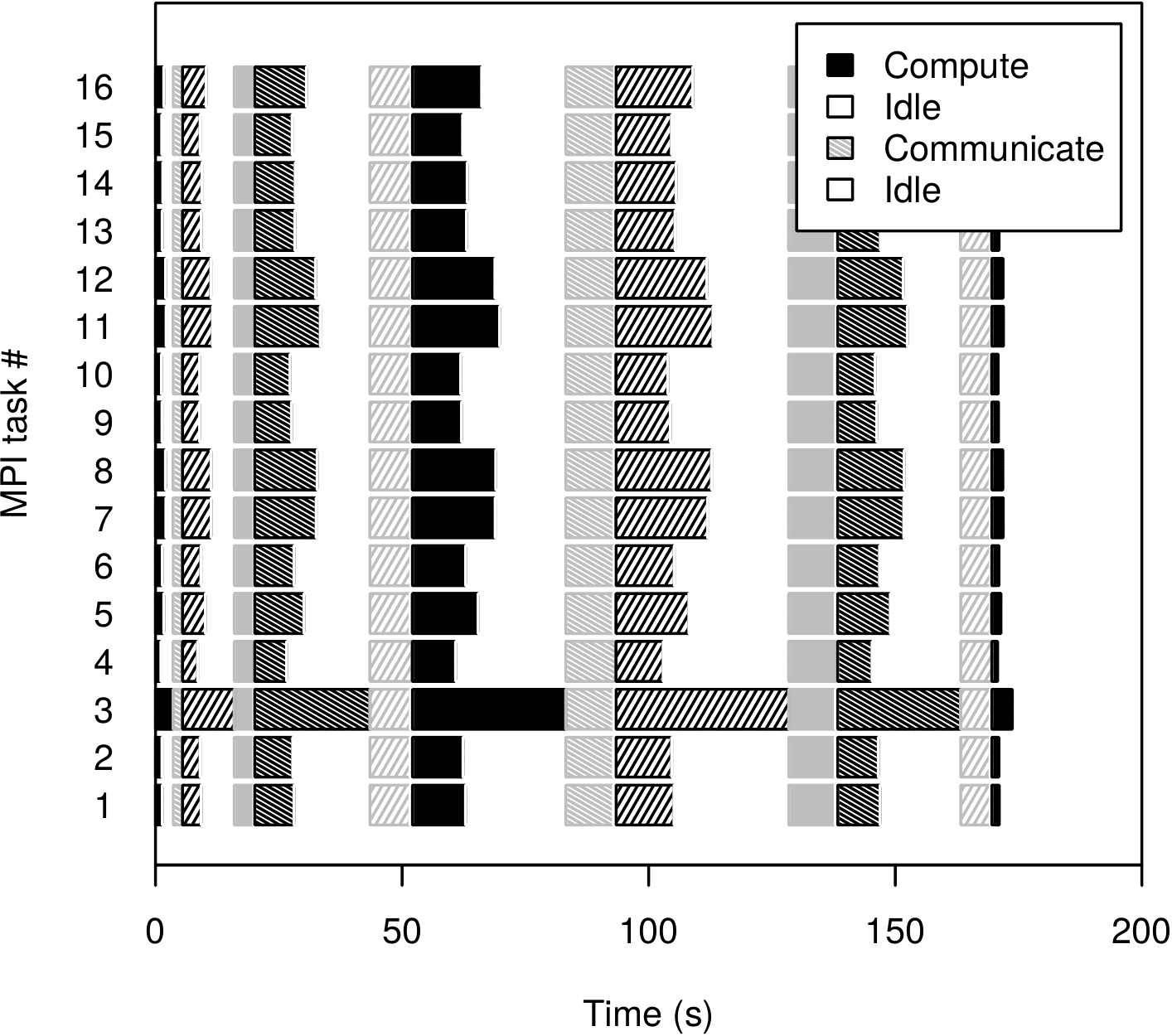}
  \includegraphics[width=0.24\textwidth]{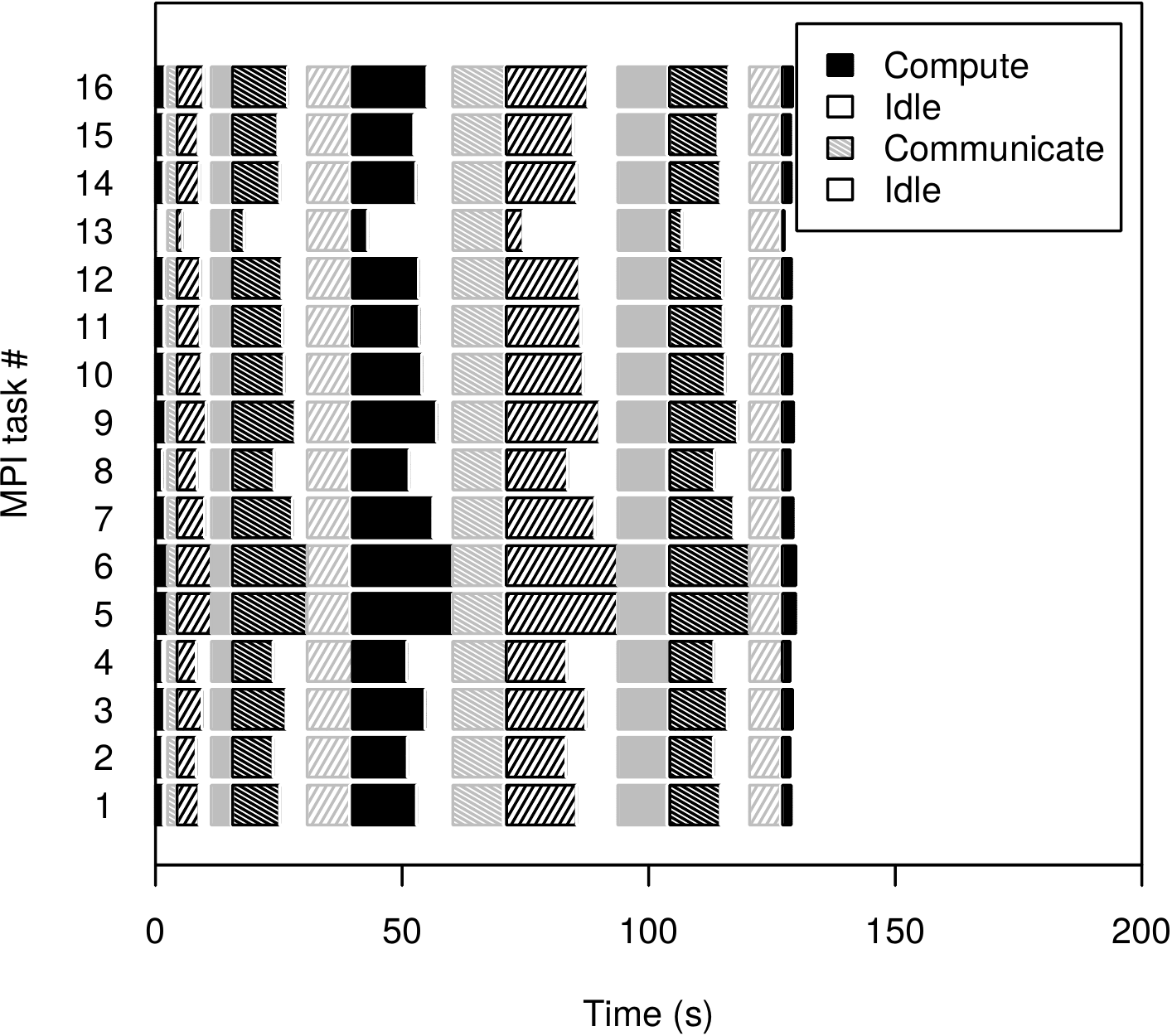}
  \includegraphics[width=0.24\textwidth]{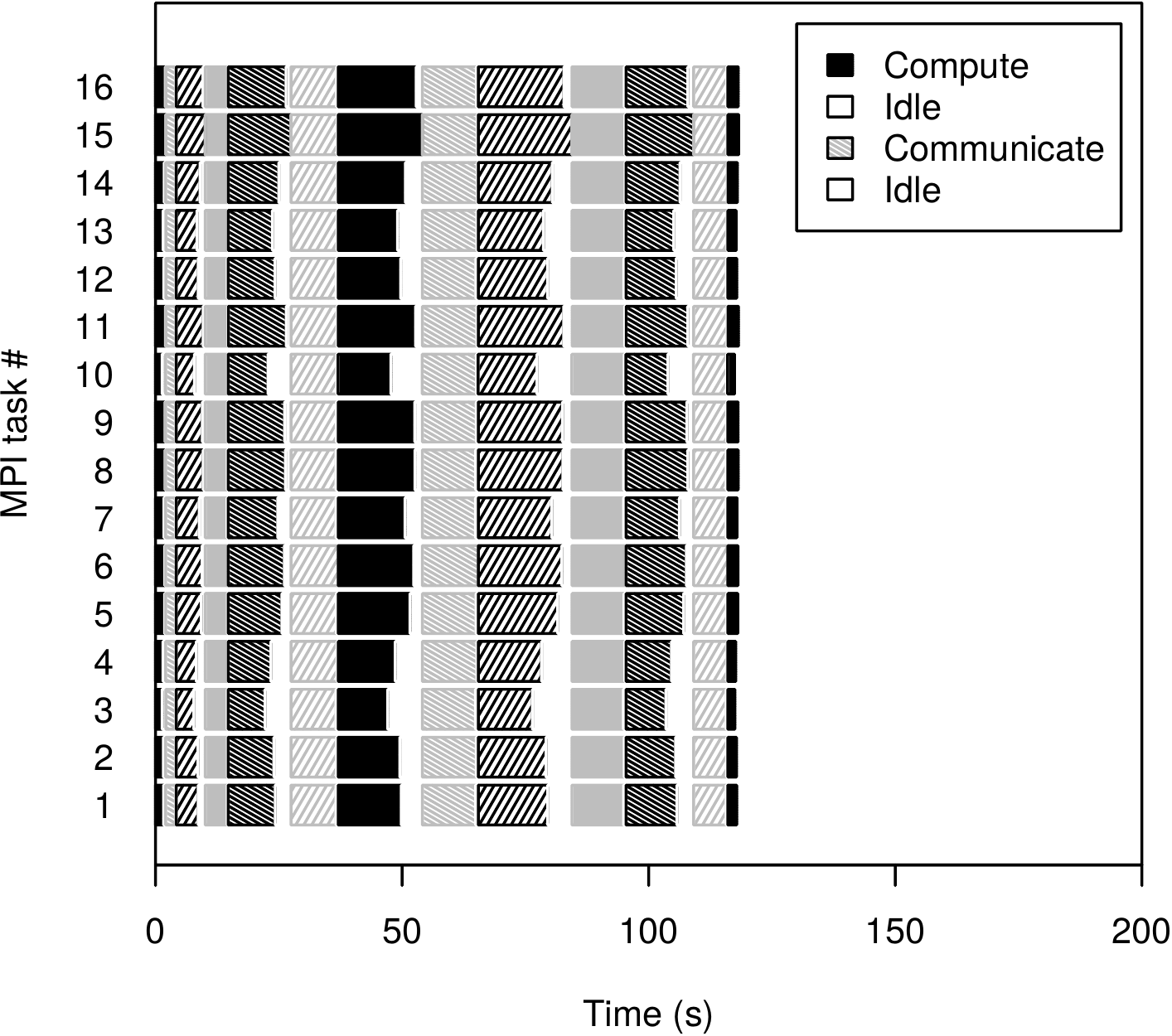}\\
  \includegraphics[width=0.24\textwidth]{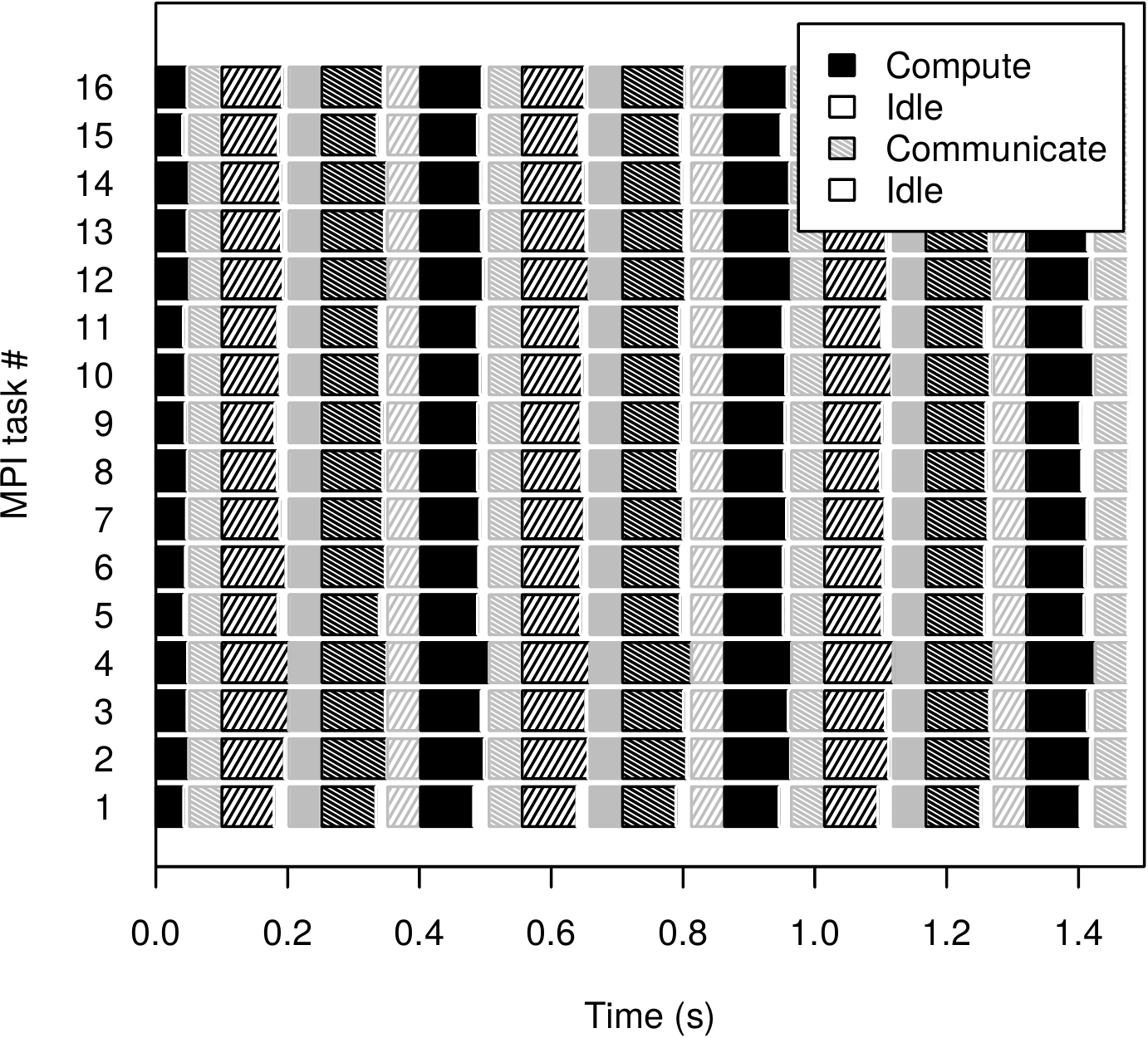}
  \includegraphics[width=0.24\textwidth]{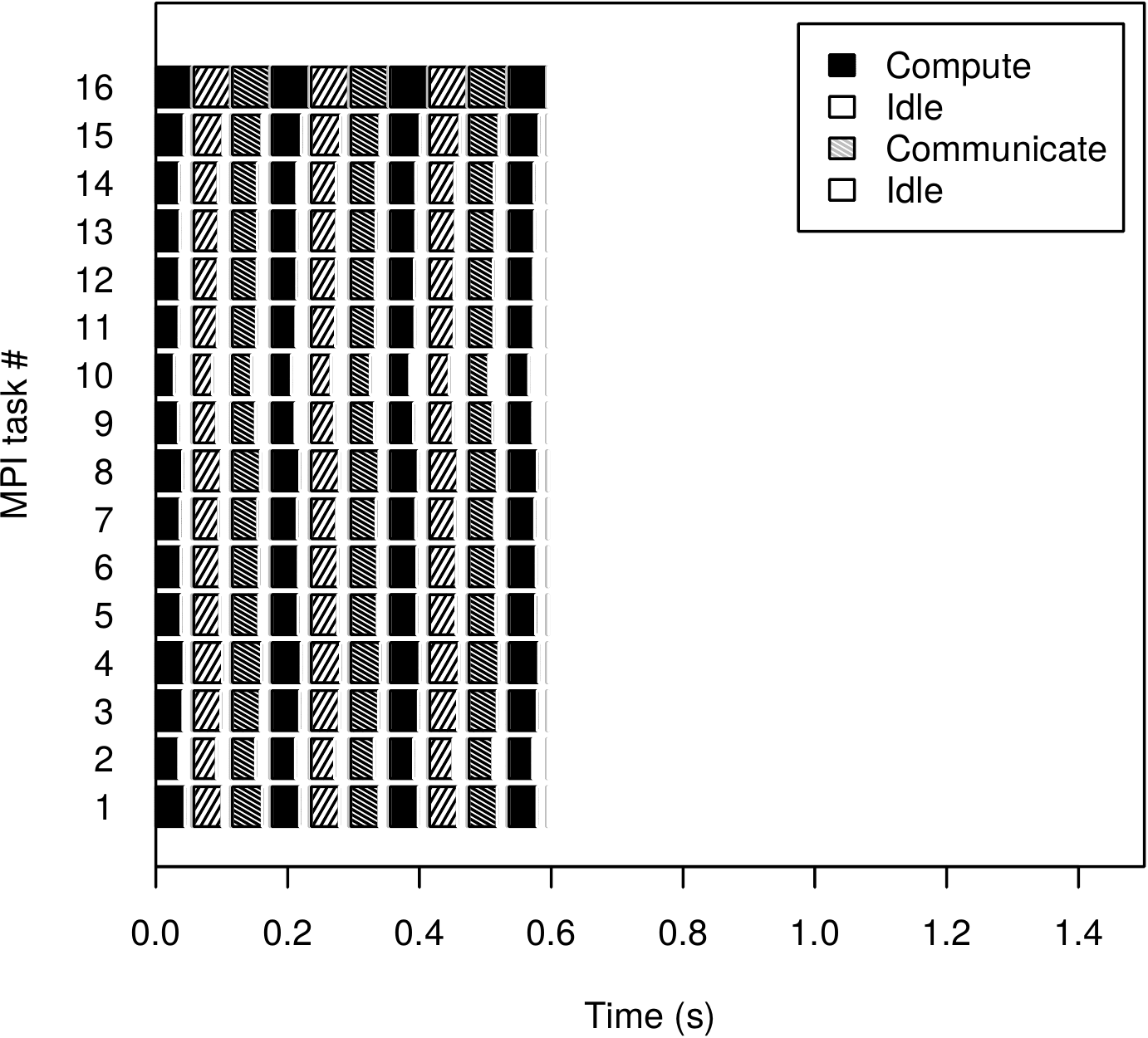}
  \includegraphics[width=0.24\textwidth]{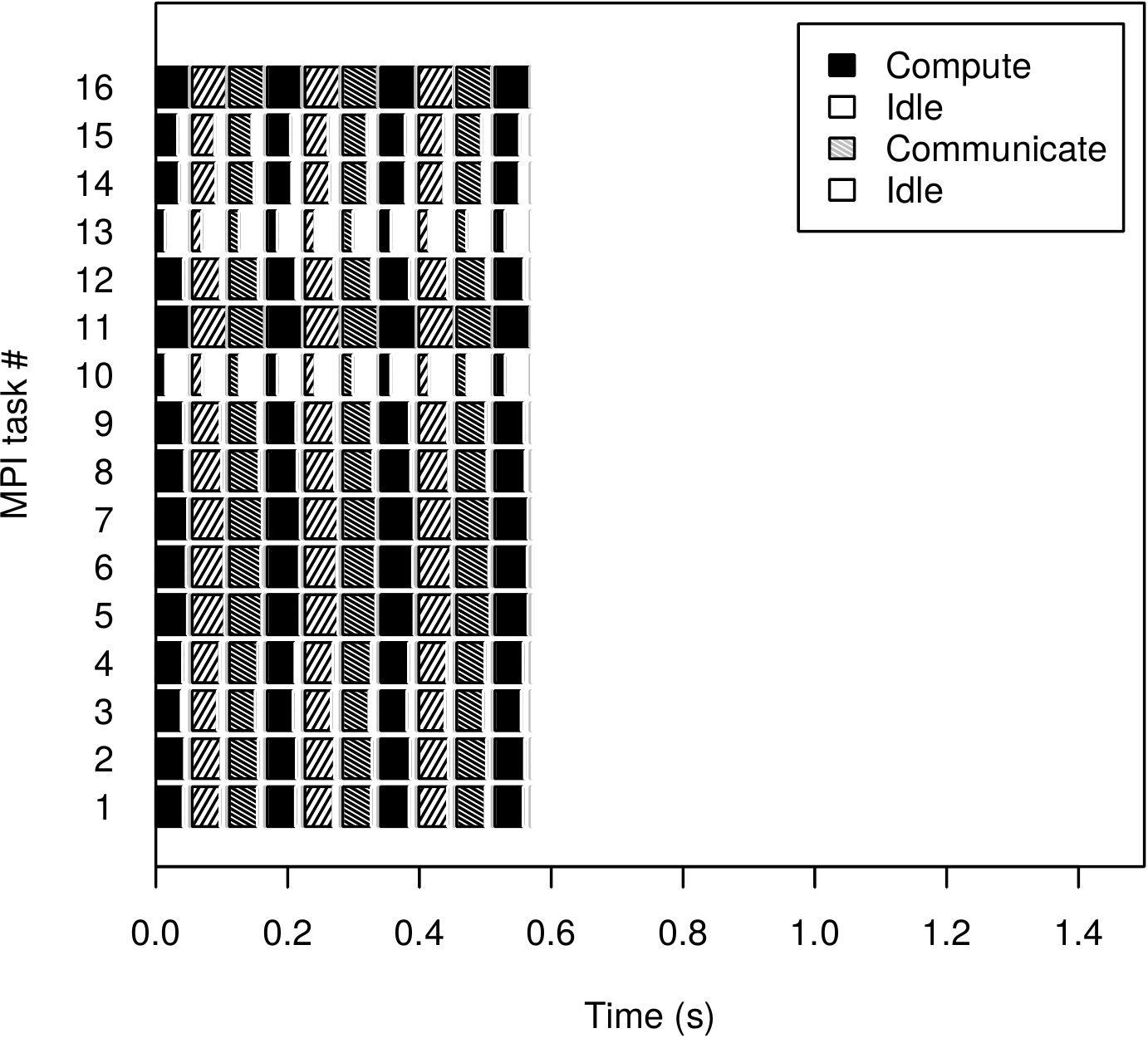}
  \includegraphics[width=0.24\textwidth]{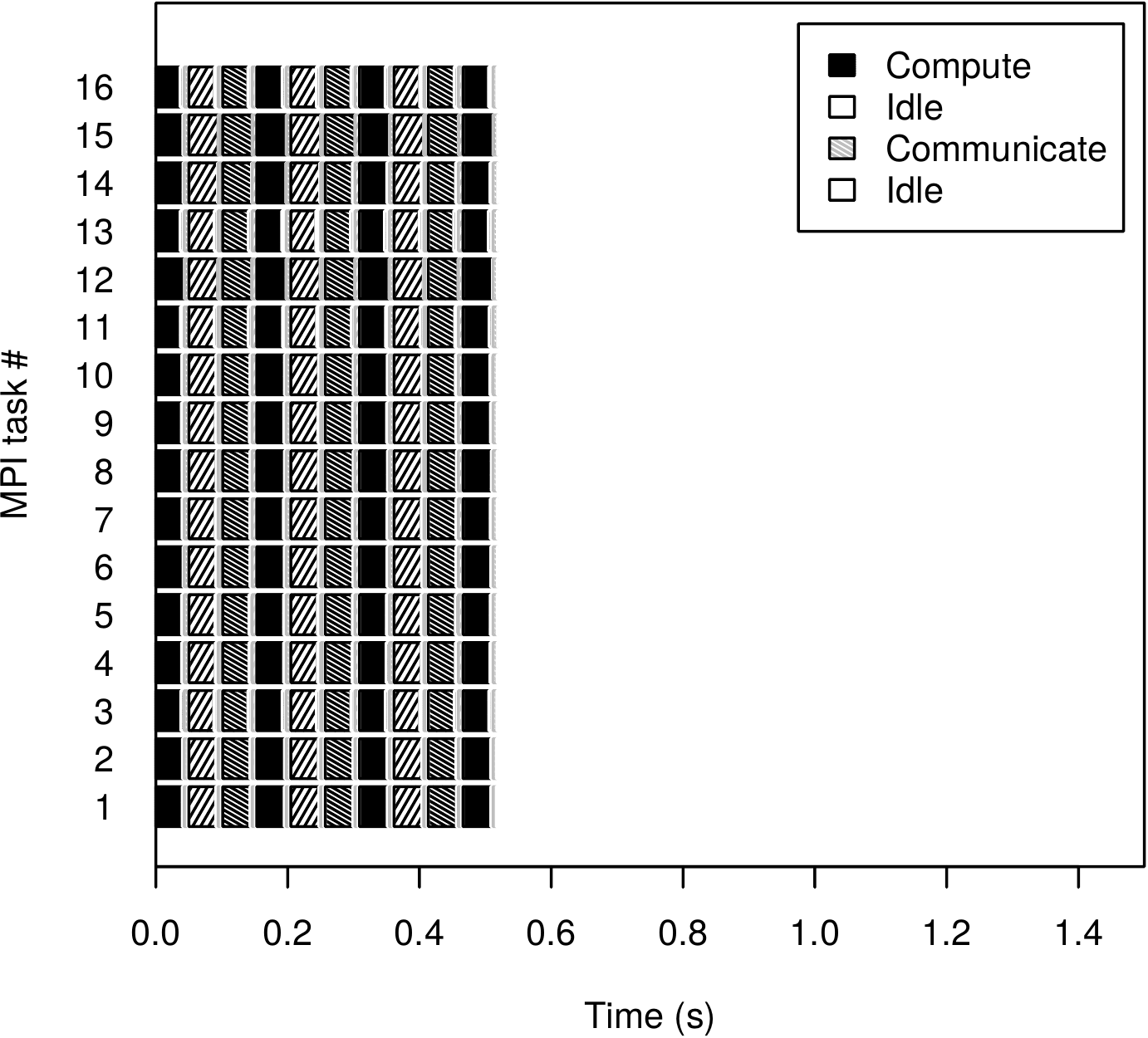}
  \caption{\verysmallfont Subgraph counting (top, single color-coding iteration with a 10-vertex template) and PageRank (bottom, 10 iterations) execution timelines on 16 tasks and 32 threads with (left to right) random, single and multi-constraint METIS, and \Pulp-MM partitioning strategies. Random ordering was used in all cases.}
  \label{fig:fascia_timeline}
\end{figure*}

To offer visual explanation of the performance of balanced constraint partitioning on total execution time, we give execution timelines in Figure~\ref{fig:fascia_timeline} of a single run of counting a 10 vertex template on the LiveJournal graph and 10 iterations of PageRank on the Webbase graph. We used the \emph{Compton} system for these tests and random, single-constraint METIS, multi-constraint METIS, and \Pulp-MM partitioning (from left to right, respectively) with random ordering. On looking at subgraph counting (top), we note first the two extreme cases. Random shows the lowest total computation times at a high cost of communication, while single objective METIS results in low communication times but high total times during the execution stages. This is due to unbalanced work among each task, which is directly proportional to the edge balance among each part. We observe that balanced multi-objective \Pulp partitioning gives the best tradeoff in terms of work balance and communication requirements. For PageRank, we note a large performance gap between Random partitioning and the other strategies. This is due to the implementation's computational and communication requirements for each task being dependent on the one hop neighborhood and per-part cut. These values are much higher with Random partitioning. We observe that \Pulp gives the best performance, due to the fact that the multiple objectives are explicitly optimizing for these metrics while keeping work balance very consistent. Overall, we notice about a 5-10\% total execution time improvement for \Pulp versus the METIS variants by using multi-objective partitioning. As noted, considering total end-to-end execution time with partitioning costs, this speedup would be even more dramatic.

\subsection{SSSP and BFS Performance}

\begin{figure*}[htb]
	\centering
	\includegraphics[width=0.95\textwidth]{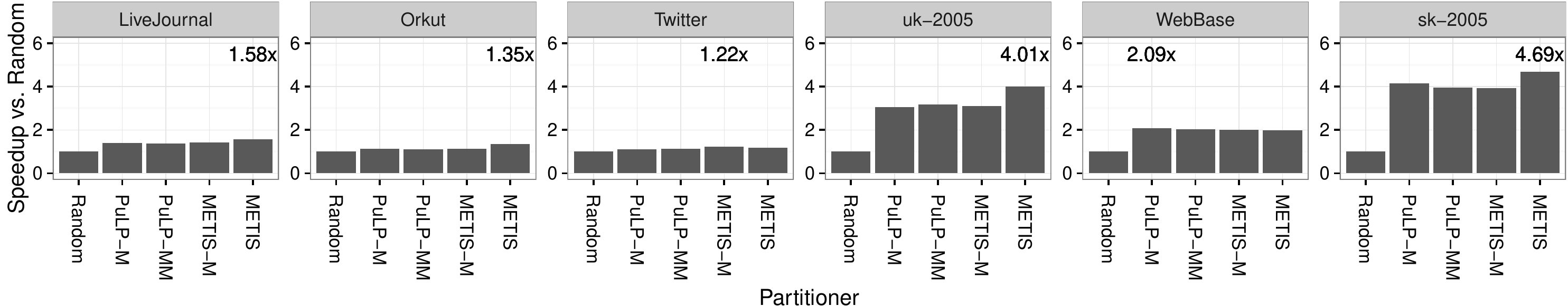}
	\includegraphics[width=0.95\textwidth]{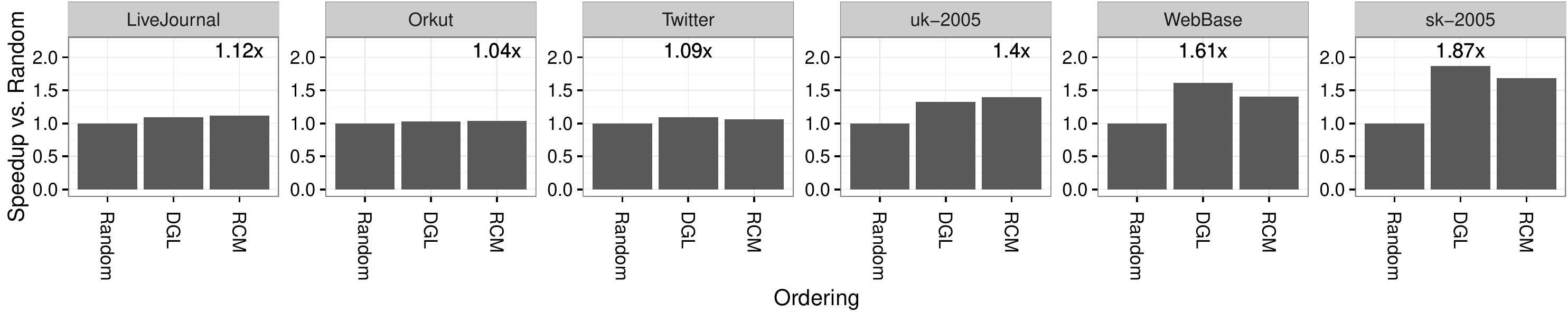}
	\caption{\verysmallfont Communication time of SSSP implementation on 64 nodes with various partitioning options (top) and computation time of SSSP with various ordering strategies (bottom).}
	\label{fig:sssp_perf}
\end{figure*}

\begin{table}[htb]
  \veryestsmallfont
  \centering
  \caption{\verysmallfont Speedups of various partitioning and ordering strategies versus random partitioning and random ordering for the SSSP counting benchmark.}\vsf
  \tabcolsep=0.1cm
  \begin{tabular}{lrrrr|rr}
    \toprule
    & \multicolumn{4}{c}{Partitioning} & \multicolumn{2}{c}{Ordering}\\
    \rb{Network}  & METIS & METIS-M & \Pulp-M & \Pulp-MM & RCM & \DGL \\
    \midrule
    LiveJournal & 1.575 & 1.422 & 1.400 & 1.372 & 1.117 & 1.097 \\
    Orkut       & 1.346 & 1.123 & 1.131 & 1.111 & 1.041 & 1.026 \\
    Twitter     & 1.172 & 1.224 & 1.109 & 1.141 & 1.063 & 1.094 \\
    uk-2005     & 4.008 & 3.122 & 3.044 & 3.187 & 1.399 & 1.328 \\
    WebBase     & 1.971 & 1.998 & 2.092 & 2.035 & 1.407 & 1.612 \\
    sk-2005     & 4.693 & 3.934 & 4.159 & 3.963 & 1.689 & 1.870 \\
    \midrule
    Overall & \textbf{2.125} & 1.907 & 1.897 & 1.884 & 1.266 & \textbf{1.304}\\
    \bottomrule
  \end{tabular}
  \label{table:sssp_perf}
\end{table}

\begin{figure*}[htb]
	\centering
	\includegraphics[width=0.95\textwidth]{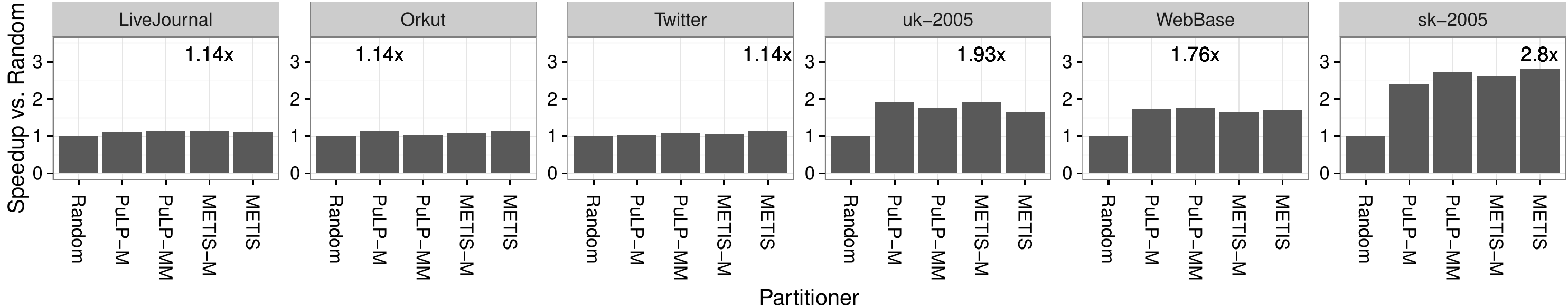}
	\includegraphics[width=0.95\textwidth]{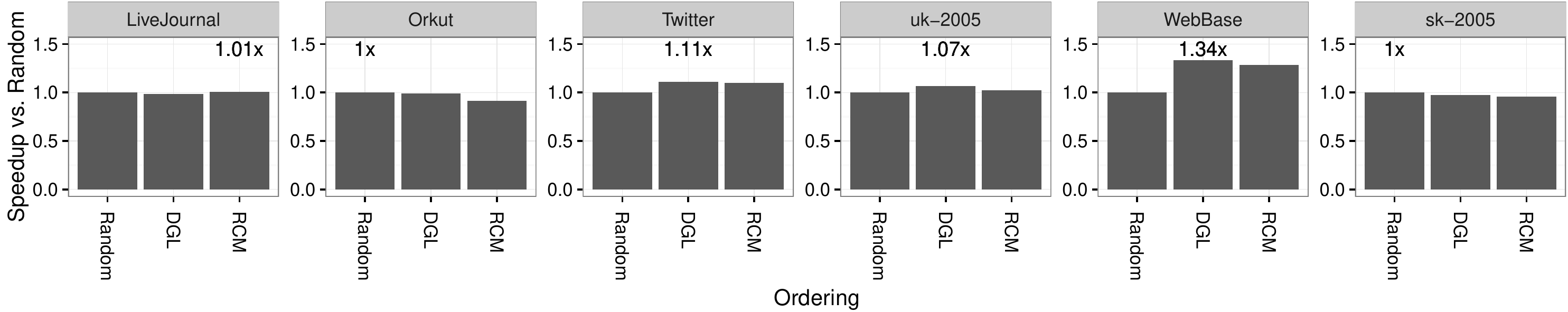}
	\caption{\verysmallfont Communication time of BFS implementation on 64 nodes with various partitioning options (top) and computation time of BFS with various ordering strategies (bottom).}
	\label{fig:bfs_perf}
\end{figure*}

\begin{table}[htb]
  \veryestsmallfont
  \centering
  \caption{\verysmallfont Speedups of various partitioning and ordering strategies versus random partitioning and random ordering for the BFS counting benchmark.}\vsf
  \tabcolsep=0.1cm
  \begin{tabular}{lrrrr|rr}
    \toprule
    & \multicolumn{4}{c}{Partitioning} & \multicolumn{2}{c}{Ordering}\\
    \rb{Network}  & METIS & METIS-M & \Pulp-M & \Pulp-MM & RCM & \DGL \\
    \midrule
    LiveJournal & 1.100 & 1.141 & 1.110 & 1.124 & 1.007 & 0.987 \\
    Orkut       & 1.125 & 1.083 & 1.138 & 1.038 & 0.913 & 0.991 \\
    Twitter     & 1.127 & 1.060 & 1.047 & 1.076 & 1.099 & 1.109 \\
    uk-2005     & 1.655 & 1.929 & 1.920 & 1.763 & 1.023 & 1.068 \\
    WebBase     & 1.709 & 1.657 & 1.732 & 1.756 & 1.287 & 1.335 \\
    sk-2005     & 2.798 & 2.624 & 2.393 & 2.737 & 0.960 & 0.977 \\
    \midrule
    Overall & \textbf{1.494} & 1.491 & 1.480 & 1.483 & 1.042 & \textbf{1.071}\\
    \bottomrule
  \end{tabular}
  \label{table:bfs_perf}
\end{table}

In this section, we analyze the performance of our SSSP and BFS implementation when using the different partitioning and ordering layouts. These benchmarks were run on 64 nodes of \emph{Blue Waters}. While the running time of distributed subgraph counting is dominated by large-scale data transfers during the communication phases, SSSP's performance is more dependent on intra-task computation, similar to PageRank, but has considerably less communication. BFS has the overall lowest communication and computation requirements out of all of the benchmarks thus far.

Figure~\ref{fig:sssp_perf} and Table~\ref{table:sssp_perf} show the speedups for communication and computation for SSSP performance with 64 MPI tasks. The top subfigure of Figure~\ref{fig:sssp_perf} shows the communication speedups relative to random partitioning for the other partitioning strategies. Due to the relatively lower communication requirements for this SSSP implementation, we correspondingly observe lower speedups relative to what was observed in Figure~\ref{fig:fascia_perf} with subgraph counting. We note that METIS gives the highest communication speedup, due to the lower overall communication load and total synchronization costs which emphasize a lower total workload than explicit balance. We observe speedups for computation times on all graphs, and especially the web crawls, with both ordering strategies. \DGL ordering gives around 30\% speedup overall.

The two subfigures of Figure~\ref{fig:bfs_perf} and Table~\ref{table:bfs_perf} give the speedup in communication time with different partitioners and speedups in computation time with different orderings for the BFS implementation. We notice similar trends to SSSP in these plots. Overall, the lower total computation and communication workload of BFS contributes to lower speedups when using better ordering and partitioning strategies compared to random.

\subsection{SPARQL Query Processing}

In this final section, we study the impact of partitioning and ordering on the performance of RDF stores and SPARQL querying, a benchmark algorithm that is very different than the previous ones. In Table~\ref{table:RDFRepRatio}, we report replication ratios observed when an undirected 2-hop guarantee is enforced. Our RDF3X-MPI implementation uses a 2-hop guarantee to partition the graph, and a lower replication ratio indicates a smaller index size, which should translate to faster query times in practice. Table~\ref{table:RDFRepRatio} compares \Pulp-MM with METIS-M and random partitioning for 16 and 64 parts on the 3 RDF graphs. Out of the 6 total graph-part count scenarios, the \Pulp-MM approach shows the lowest replication ratio for half of them. Note that none of these partitioners are explicitly optimizing for this metric, so the performance of \Pulp-MM in this instance is indirect.

\begin{table}[htb]
	\centering
	\caption{Distributed RDF store replication ratios using various partitioning strategies. An undirected 2-hop guarantee is enforced. Lower values are better and best value for each graph and parts count is in bold.}\vsf
	\tabcolsep=0.1cm
	\begin{tabular}{lrrr|rrr}
		\toprule
		& \multicolumn{3}{c}{16-way} & \multicolumn{3}{c}{64-way}\\
		\rb{Partitioning}  & BSBM & LUBM & DBpedia & BSBM & LUBM & DBpedia\\
		\midrule
    Random    & 7.256          & 10.58          & 5.580 
              & 22.07          & 34.84          & 10.50 \\
    METIS     & 5.566          & 9.714          & \textbf{1.552} 
              & 19.02          & 36.56          & 2.257 \\
    METIS-M   & 5.577          & 9.146          & \textbf{1.552} 
              & 19.01          & 38.02          & \textbf{2.255} \\
    \Pulp-M   & 5.308          & \textbf{8.944} & 1.905 
              & 14.73          & 36.86          & 2.815 \\
    \Pulp-MM  & \textbf{5.112} & 9.227          & 2.448
              & \textbf{13.78} & \textbf{29.94} & 2.963 \\
		\bottomrule
	\end{tabular}
	\label{table:RDFRepRatio}
\end{table}

In Table~\ref{table:query_time}, we report sum of query times of RDF3X-MPI averaged over the BSBM, LUBM, and DBpedia data sets. We use a selection of queries from the Berlin SPARQL Benchmark. We use the 16 part partitions for this test and additionally look at the performance affects of the three ordering strategies. \Pulp-MM partitioning with random ordering yields the best performance, while \Pulp-MM further demonstrates the highest performance when using the other two ordering strategies as well. This corresponds to \Pulp-MM having the lowest replication ratios. We note that since \Pulp-MM is faster and much more memory-efficient than METIS, this is a promising result, and future work can attempt to optimize \Pulp for the one and two hops replication ratio metrics for further improvements. The effect of ordering strategy on query times is interesting, in that the higher-locality orderings demonstrate correspondingly worse performance. To store the RDF data, RDF3X-MPI converts the input RDF graph structure into multiple indexes, which are created by sorting the RDF data, creating B+ trees, and then performing compression. We note that the worsened performance with locality-optimized ordering is most likely an artifact of this pre-processing stage. This further indicates that knowledge of a graph analytic's algorithmic details is important when determining an optimal graph layout, as unexpected and counter-intuitive performance impacts are a real possibility.


\begin{table}[htb]
	\verysmallfont
	\centering
	\caption{Total query times in seconds relative for the various partitioning and ordering strategies, summed over all 3 graphs with 16 parts.}\vsf
	\tabcolsep=0.1cm
	\begin{tabular}{lrrr}
		\toprule
		& \multicolumn{3}{c}{Ordering}\\
		\rb{Partitioning}  & Random & \DGL & RCM\\ 
		\midrule
		Random    & 3.41          &  4.58 &  4.32 \\
		\Pulp-M   & 3.41          &  3.97 &  3.94 \\
		\Pulp-MM  & \textbf{3.32} &  4.01 &  3.51 \\
    METIS     & 3.71          &  4.41 &  3.91 \\
		METIS-M   & 3.87          &  4.20 &  4.13 \\
		\bottomrule
	\end{tabular}
	\label{table:query_time}
\end{table}




\section{Related Work}

We selected METIS and RCM for comparison to the partitioning and ordering aspects of \DGL, as they represent the most popular and current state-of-the-art approaches for these problems in terms of both speed of computation and quality produced. There are various other partitioning algorithms and methods, including multi-level partitioners similar to METIS~\cite{kahip, chaco, dimacs}, coordinate and geometry-based partitioners~\cite{zoltan2}, and hypergraph partitioners~\cite{patoh}. Hypergraph partitioners can often calculate higher quality partitions than graph partitioners for regular matrices, but at a considerably higher cost to compute. Other graph partitioners have utilized label propagation in single or multilevel approaches~\cite{xdgp,UG13,partbillion,kahipPar}, demonstrating improved algorithm execution times with these partitions versus \naive methods. 

However, while some of these partitioners produce very high partition quality with good computational efficiency~\cite{kahipPar}, they only consider single constraint partitioning scenarios. As we've demonstrated, using multi-constraint partitioning is important for optimal algorithm performance with high computation loads. Though we acknowledge that the tradeoff of pre-processing time for runtime performance with communication-bound algorithms is going to be application-specific. Other recent work~\cite{deveci2015fast} has correspondingly demonstrated that complex partitioning scenarios beyond single objective partitioning optimizing for edge cut and/or communication volume is a necessary consideration for optimal performance in other distributed computations.  

Recently, other partitioning methods have been developed with goals similar to that of \Pulp, in terms of striking the balance between both high scalability and high quality of part computation. We compare to the most notable of these, FENNEL~\cite{fennel}, using their published results. On a large Twitter dataset, FENNEL reports an edge cut relative to METIS of 0.56$\times$, 1.19$\times$, and 1.33$\times$ (lower is better) for computing 2, 4, and 8 parts, respectively. However, the relative imbalance for FENNEL is higher than the balance constraint imposed on METIS, which makes a lower edge cut considerably easier to achieve. Using the same graph, we impose equivalent balance constraints for \Pulp and METIS and compute relative edge cuts of 1.19$\times$, 1.05$\times$, and 1.13$\times$ again for 2, 4, and 8 parts. FENNEL reports a time of 40 minutes to partition the Twitter graph. On this graph, which is smaller than the Twitter graph we used in our primary results, we compute these partitions with \Pulp in about 5 minutes. For 16 parts on LiveJournal, use of FENNEL speeds up PageRank execution relative to random (hash) partitioning by about 1.18$\times$. On the same graph and number of parts, we reported an improvement of 2.83$\times$ with \Pulp relative to random partitioning.

In addition to RCM ordering, Cuthill-McKee (CM)~\cite{CMalg}, nested dissection~\cite{METIScode}, and Approximate Minimum Degree (AMD)~\cite{AMDalg} are a few examples of sparse matrix reordering strategies used in the past. Some techniques, such as space-filling curves~\cite{spacecurve} or spectral bisection and orderings based on calculated eigenvectors~\cite{spectraleigen} have been utilized for both partitioning and ordering of sparse matrices. 
Ordering methods on irregular networks such as social and Internet graphs has been studied for the purposes of visualization~\cite{orderviz} and compression~\cite{shingle,llp}. Although the authors know of no performance analysis of the effects of applying these ordering techniques to distributed graph computation in literature, promising future work might involve utilizing and optimizes these ordering for such purposes.

We omit direct comparison using the distributed-memory graph processing frameworks mentioned previous~\cite{graphLab, PowerGraph, PowerLyra, Giraph, Trinity, PEGASUS}, as prior work has demonstrated a several orders-of-magnitude performance difference between them and optimized code~\cite{ipdps16, cost}. However, we acknowledge the goal of these frameworks is often programmer efficiency rather than pure performance. Implementing our methods within such frameworks would make for interesting future work.

\section{Conclusions}

In this paper, we present a methodology for distributed \emph{graph layout} (partitioning, vertex ordering). The partitioning method for our layout, \Pulp, is based on the scalable label propagation community detection method. The partitions produced are comparable in quality to the $k$-way multilevel partitioning scheme in METIS, but only take a fraction of the execution time. Our \DGL vertex ordering strategy can also improve computational performance of graph computations that consist of a high proportion of irregular accesses. 

Additionally, we give a comprehensive performance analysis by examining the effect of graphs layouts on a graph analytics workload. In general, we note that graph analytics which have a high relative computation cost can benefit greatly from a locality-optimizing vertex ordering strategy. Graphs analytics that have a relatively high communication volume but few synchronizations might benefit most from a partitioning that optimizes explicitly for edge cut, while computations which consist of numerous synchronizations would benefit from a more balanced partitioning in terms of per-task communication loads. In general, the higher the computation to communication ratio for an analytic, the greater the impact of partitioning and ordering.



\ifCLASSOPTIONcompsoc
 \section*{Acknowledgments}
\else
 \section*{Acknowledgment}
\fi

This research is part of the Blue Waters sustained-petascale computing project, which is supported by the National Science Foundation (awards OCI-0725070, ACI-1238993, and ACI-1444747) and the state of Illinois. Blue Waters is a joint effort of the University of Illinois at Urbana-Champaign and its National Center for Supercomputing Applications. This work is also supported by NSF grants ACI-1253881, CCF-1439057, and the DOE Office of Science through the FASTMath SciDAC Institute. Sandia National Laboratories is a multi-program laboratory managed and operated by Sandia Corporation, a wholly owned subsidiary of Lockheed Martin Corporation, for the U.S. Department of Energy's National Nuclear Security Administration under contract DE-AC04-94AL85000.

\ifCLASSOPTIONcaptionsoff
  \newpage
\fi



\bibliographystyle{IEEEtranS}
\bibliography{graphpart}

\end{document}